\documentclass[twocolumn,apj]{openjournal}

\usepackage[breaklinks,colorlinks,citecolor=blue,urlcolor=blue]{hyperref}
\usepackage{amsmath}

\usepackage[T1]{fontenc}
\usepackage{newtxtext,newtxmath}
\usepackage{graphicx}
\usepackage{amsfonts,ulem}
\usepackage{soul}
\usepackage{hyperref}
\usepackage{enumitem}
\usepackage[hyphens]{xurl}
\usepackage{multirow}

\newcommand{\orcidauthor}[3]{\author{\href{http://orcid.org/#1}{#2$^{#3}$}}}

\renewcommand{\arraystretch}{1.2}  

\shorttitle{{\tt HoliGRALE}}
\shortauthors{Perera et al.}

\begin{document}

\title{\vspace{-0.8cm}(Monty Python and the) {\tt HoliGRALE}: A Hybrid {\tt GRALE} Lens Inversion Methodology Diagnostically Evaluated with Synthetic and Real Data \vspace{-1.5cm}}

\orcidauthor{0000-0002-4693-0700}{Derek Perera}{1,^\dagger}
\orcidauthor{0000-0002-3648-8031}{Jori Liesenborgs}{2,^\star}
\orcidauthor{0000-0002-2901-5260}{John H. Miller Jr.}{1} 
\orcidauthor{0009-0006-5681-6297}{Ashley Francis}{1}
\orcidauthor{0000-0002-6039-8706}{Liliya L. R. Williams}{1,^\star}
\orcidauthor{0000-0003-0203-3853}{Birendra Dhanasingham}{1}
\orcidauthor{0000-0001-9065-3926}{Jose M. Diego}{3}

\affiliation{$^1$School of Physics and Astronomy, University of Minnesota, Minneapolis, MN, 55455, USA}
\affiliation{$^{2}$UHasselt – Flanders Make, Digital Future Lab, Wetenschapspark 2, B-3590, Diepenbeek, Belgium.}
\affiliation{$^{3}$Instituto de Física de Cantabria (CSIC-UC). Avda. Los Castros s/n. 39005 Santander, Spain.}

\thanks{\vspace{0.1cm}$^\dagger$Corresponding author: \href{mailto:perer030@umn.edu}{perer030@umn.edu}}
\thanks{\vspace{0.1cm}$^\star$E-mail: \href{mailto:jori.liesenborgs@uhasselt.be}{jori.liesenborgs@uhasselt.be} (J.L.); \href{mailto:llrw@umn.edu}{llrw@umn.edu} (L.L.R.W.)}

\begin{abstract}\vspace{0.2cm}
    We present {\tt HoliGRALE} (Hybrid, only light included, GRAvitational LEnsing), a hybrid lens modeling extension of the freeform method {\tt GRALE}. {\tt HoliGRALE} incorporates physically motivated parametric components within the freeform genetic algorithm {\tt GRALE}. Using three synthetic clusters from IllustrisTNG, each of varying complexity, we evaluate its performance according to its accuracy in reconstructing the density distribution, image positions, magnifications, and time delays. We find that {\tt HoliGRALE} generally outperforms {\tt GRALE} on all fronts, and shows greater accuracy in regions containing a high image number density. We estimate the error budget in projected density to be $<5\%$ for simple, few component clusters, and $\sim$5--10\% for complex, multi-component clusters in the regions constrained by observed images. We then use {\tt HoliGRALE} to model three lenses: RX\,J2129.7$+$0005, MACS\,J0138.0$-$2155, and Abell S1063. In each case, {\tt HoliGRALE} is able to accurately reconstruct observed magnifications and time delays. In its predictions, {\tt HoliGRALE} is found to be more precise than {\tt GRALE}.  We particularly highlight the predicted time delay for the next arriving image of SN\,Requiem to be $4346.61^{+490.27}_{-818.36}$ days ($\approx$ April 2026 -- October 2029), consistent with previous work, albeit with less precision. In Abell S1063, we report the reconstruction of a main dark matter halo core radius of $89.9\pm33.2$ kpc, in excellent agreement with past work and useful in testing theories of dark matter. Our results emphasize the expected error budget of {\tt HoliGRALE}, which should be mentioned and scrutinized in future applications of {\tt HoliGRALE} for other lens systems.
\end{abstract}

\section{Introduction}

The accurate modeling of cluster-scale gravitational lenses is essential for numerous areas of astrophysics and cosmology, including, but not limited to, constraining star formation history at high redshifts \citep[e.g.,][]{meena25,li25,zackrisson26}, characterizing high redshift stellar transients and supernovae (SNe) \citep[e.g.,][]{kelly22,diego26a,dhanasingham26,williams26}, constraining the Hubble constant ($H_0$) \citep[e.g.,][]{napier23,kelly23,martinezarrizabalaga24,pascale25,pierel26}, and constraining properties of dark matter \citep[e.g.,][]{dai18,williams24,broadhurst25,perera26}. The primary challenge in cluster lens modeling, therefore, is ascertaining a detailed understanding of the lensing degeneracies and systematics associated with lens models, as these can have adverse effects on the constraints obtained from them \citep[see e.g.,][]{liesenborgs08a,liesenborgs08b,liesenborgs12,williams19,lasko23,li25,limousin25}. To address this, diagnostic tests are necessary to evaluate the error budget and systematics of lens modeling techniques \citep{ghosh20,cha26}.

The timeliness of this issue needs to be appreciated; observations with the James Webb Space Telescope (JWST) are increasing the number of multiply imaged systems in cluster lenses by orders of magnitude, and data from numerous recently completed surveys are becoming publicly available, such as VENUS \citep[Cycle 4, GO-6882, PIs: S. Fujimoto \& D. Coe; ][]{fujimoto25} and SLICE \citep[Cycle 3, GO-5594, PI; G. Mahler; ][]{cerny26}. It is necessary to critically and rigorously evaluate lens modeling techniques and paradigms in order to understand the associated uncertainty budgets. A common way to do this has been in meta studies comparing different lens modeling procedures \citep{priewe17,remolina18,raney20a,perera25b}. In general, these works find that lens models agree on large scales, however, exhibit more discrepancies on smaller scales. \cite{perera25b} quantified the level of disagreement between the mass reconstructions of various lens models to be $\sim9\%$. 

Probing systematics of lens models is best done in a data challenge using mock cluster lenses. The most influential study like this is \cite{meneghetti17}, which compared the accuracy in the reconstruction of two simulated cluster lenses across multiple modeling algorithms. Their mock cluster lenses, Ares and Hera, have been used by the lens modeling community to validate many subsequent new lens modeling algorithms \citep{cha23,cha26}. There are numerous other ways to test systematics, such as by comparing the effects of input image positions \citep{acebron17,sebesta19,zimmerman21}, evaluating the impact of model priors \citep{raney21,lasko23}, employing a jackknifing procedure to input image positions \citep{nishida25}, considering extended arcs \citep{eid25}, and considering line-of-sight effects \citep{williams18,raney20b,madhava24,roche26}.

There are two main classes of cluster lens models: parametric and freeform. Parametric models generally use observationally motivated priors for cluster components, such as cluster member galaxies, the brightest cluster galaxy (BCG), and the cluster-scale dark matter halo. These methods are physically motivated, but may be biased, unable to recover small scale substructures, and underestimate uncertainties \citep{priewe17}. Freeform models only use lensed image positions as input, and thus offer an unbiased and flexible view of the lens. However, this flexibility can be a disadvantage, as freeform models can be less precise, predict physically disfavored mass components, and be susceptible to overfitting in low number of image systems. A middle ground approach between the two classes are hybrid models, which was a suggested approach from the results of \cite{meneghetti17}. Hybrid models incorporate parametric constraints and priors within a freeform modeling framework, thereby preserving the flexibility of freeform modeling and the physical realism of parametric modeling.

In this work, we develop {\tt HoliGRALE} (Hybrid, only light included, GRAvitational LEnsing), a hybrid lens modeling paradigm and extension of the existing freeform {\tt GRALE} \citep{liesenborgs06,liesenborgs07,liesenborgs20}. {\tt HoliGRALE} incorporates parametric basis functions within the freeform {\tt GRALE} genetic algorithm that represent cluster member galaxies. While {\tt GRALE} has been used extensively to model cluster lenses \citep{liesenborgs08b,liesenborgs09,mohammed14,mohammed15,sebesta16,sebesta19,ghosh21,ghosh23,perera24,perera25a,limousin25}, {\tt HoliGRALE} has only been used in a limited capacity in previous works \citep{liesenborgs20,perera24,liesenborgs24,perera25a}. It has already found early success in being found to be the optimal model to reconstruct the observed transient detection rate \citep{li25}. As hybrid models become more popular, it is necessary to stress test {\tt HoliGRALE} in order to understand and quantify its systematic uncertainties. 

We evaluate the performance of {\tt HoliGRALE} in terms of its accuracy in reconstructing the density distribution, image positions, magnifications, and time delays of cluster lenses. To validate {\tt HoliGRALE}, we perform a data challenge using three synthetic clusters of varying complexity from the TNG-Cluster simulation from the IllustrisTNG project \citep{nelson19}. This approach is meant to replicate a wide range of cluster lens scenarios, something not seen in systematics studies, which historically have adopted a single mock lens for validation. We then apply {\tt HoliGRALE} to three real cluster lenses with different observed properties that are of current interest. The ultimate goal of this work is to establish the baseline expectation for the uncertainty budget and systematics of {\tt HoliGRALE} such that its results, when used in the future, can be interpreted properly.

We describe the synthetic cluster properties in Section \ref{txt:clusters}. We describe the algorithms of {\tt GRALE} and {\tt HoliGRALE} in Section \ref{txt:holigrale}. We define the metrics we employ in Section \ref{txt:metrics}. Lastly, we evaluate the performance of {\tt HoliGRALE} on the synthetic clusters and real lenses in Sections \ref{txt:synthetic} and \ref{txt:real}, respectively.

Throughout this paper we assume a cosmology of $H_0=67.74$ km s$^{-1}$ Mpc$^{-1}$, $\Omega_{\rm m,0}=0.3089$, and $\Omega_\Lambda=0.6911$ \citep{planck18} unless otherwise specified. We define the commonly used notation of gravitational lensing formalism in Table \ref{tab:lensing}.

\begin{table}
\centering
\caption{Gravitational Lensing Quantities}
\label{tab:lensing}
\setlength{\tabcolsep}{2.45pt}
\begin{tabular}{@{}ccc@{}}
\hline
Symbol & Description & Definition / Units \\
\hline
$\Sigma$ & Surface Mass Density & [$M_{\odot}/{\rm arcsec}^2$] \\
$\kappa$ & Convergence & $\Sigma/\Sigma_{\rm crit}$ \\
$\boldsymbol{\theta}$ & Angular Position & Angular units\\
$\boldsymbol{\alpha}(\boldsymbol{\theta})$ & Deflection angle & $\frac{1}{\pi}\int\kappa(\boldsymbol{\theta}')\frac{\boldsymbol{\theta}-\boldsymbol{\theta}'}{|\boldsymbol{\theta}-\boldsymbol{\theta}'|^2}d^2\boldsymbol{\theta}'$ \\
$\psi(\boldsymbol{\theta})$ & Lens Potential & $\frac{1}{\pi}\int\kappa(\boldsymbol{\theta}')\ln|\boldsymbol{\theta}-\boldsymbol{\theta}'|d^2\boldsymbol{\theta}'$ \\
$\boldsymbol{\beta}$ & Source Position & $\boldsymbol{\theta}-\boldsymbol{\alpha}(\boldsymbol{\theta})$\\
$\mu$ & Magnification & $1/\det\left(\partial\boldsymbol{\beta}/\partial\boldsymbol{\theta}\right)$ \\
$\tau$ & Time Delay surface & $\propto\frac{1}{2}\left(\boldsymbol{\theta}-\boldsymbol{\beta}\right)^2 - \psi(\boldsymbol{\theta})$\\
\hline
\end{tabular}\\
\medskip{Definitions of the standard quantities of gravitational lensing. }
\end{table}

\section{Synthetic Clusters}\label{txt:clusters}

\begin{figure}
\includegraphics[trim={5.2cm 0cm 5.0cm 0.1cm},clip,width=0.49  \textwidth]{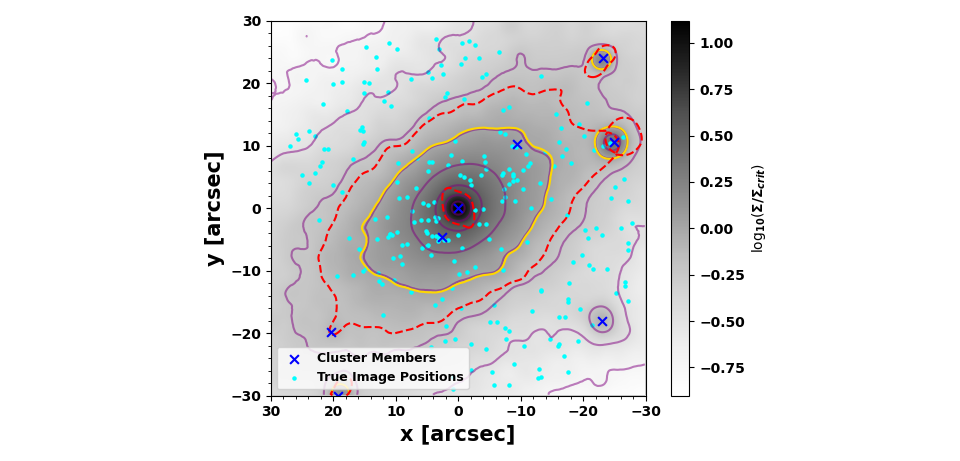}
\includegraphics[trim={5.2cm 0cm 5.0cm 0.1cm},clip,width=0.49\textwidth]{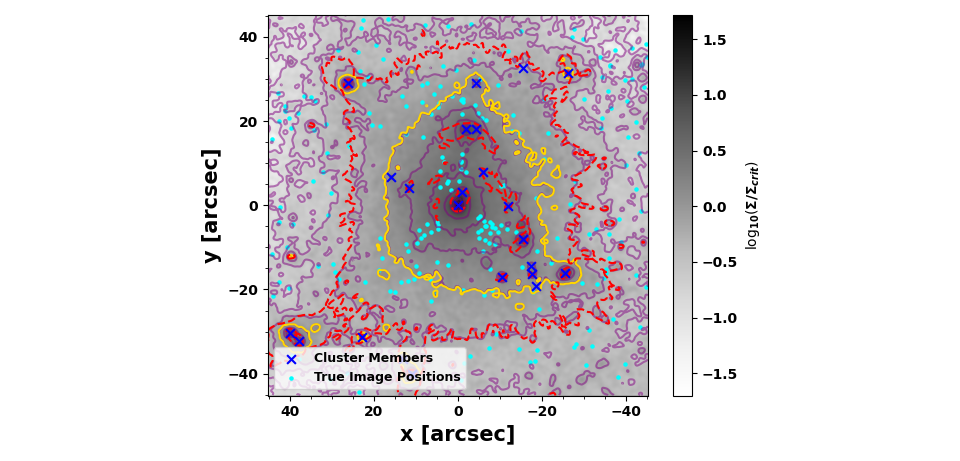}
\includegraphics[trim={5.2cm 0cm 5.0cm 0.1cm},clip,width=0.49\textwidth]{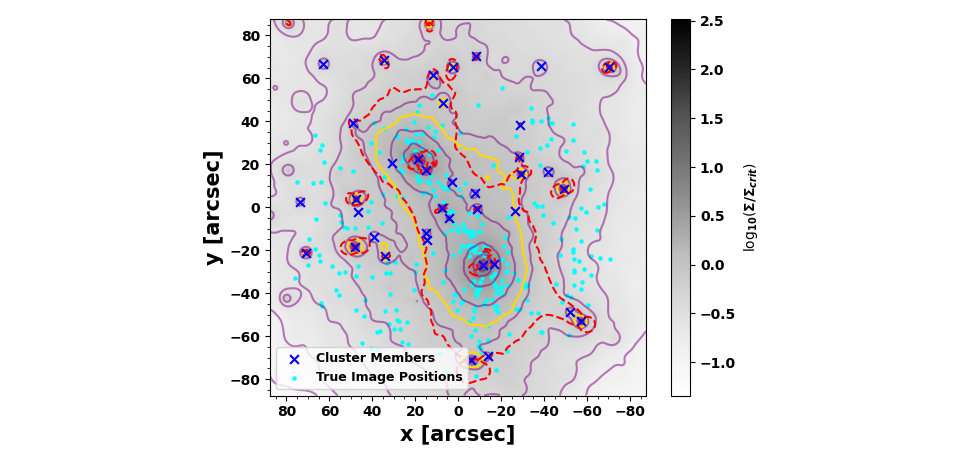}
\caption{The three synthetic clusters from IllustrisTNG used in this work. The top, middle, and bottom panels depict the true $\kappa$ distributions for Harriet (Section \ref{txt:c0}), Maka Ska (Section \ref{txt:c1}), and Nokomis (Section \ref{txt:c2}), respectively. The red dashed line indicates the critical curve for $z=3$. The solid gold line indicates $\kappa=1$. Blue crosses and cyan dots show cluster member galaxies and the true image positions for the set of 100 sources, respectively. Purple contours trace out $\kappa$ in spacings of $\log \kappa = 0.25$.}
\label{fig:synthclusters}
\end{figure}

Figure \ref{fig:synthclusters} shows the density distributions for the three synthetic clusters used in this analysis. Each is derived from TNG-Cluster, a large volume cosmological hydrodynamical simulation from the IllustrisTNG project \citep{nelson19,nelson24}. Like all IllustrisTNG simulations, TNG-Cluster uses the moving-mesh code AREPO and assumes a Planck Collaboration XIII cosmology \citep{springel10,planck16}. With a volume of roughly 1 Gpc$^3$ and DM particle mass of $6.1 \times 10^7 M_{\odot}$, TNG-Cluster samples the most massive DM halos ($M_{200}> 10^{14.5} M_{\odot}$) with excellent resolution, which is why it is chosen for this work. We note that all three synthetic clusters in this work are similar to Hera \citep{meneghetti17} in that they are derived from simulations. We opt to utilize IllustrisTNG clusters for this work in order to validate {\tt HoliGRALE} across multiple different realizations of cluster complexity. Further specific details for each cluster are given in the following sections. For easy reference, we name each cluster accordingly in the subsection title\footnote{In the most recent {\tt GRALE} diagnostic paper, synthetic clusters were named ``Irtysh'' after a river in central Asia \citep{ghosh20}. Keeping with the theme of bodies of water, we name the three synthetic clusters in this paper after lakes nearby to the University of Minnesota.}.

The mock images for each synthetic cluster were generated by first sampling a source redshift, $z_s$, from a Gaussian with $\mu_{z_{\rm s}} = 2.5$, $\sigma_{z_{\rm s}}=1$, and $z_{{\rm s, max}} = 6.5$. With the re-scaled lens potential, random source positions are uniformly sampled within the modeling window and forward modeled by differentiating and solving the lens equation. A joint magnification-location selection effect is applied to the resulting mock images, rejecting lower magnification images as they lie nearer to galaxy centers, depending on the approximate size of the galaxy.

The selection effect is applied to images when source redshifts and positions result in more than one image lying within the modeling window. Typically, central images do not pass the selection criteria. If the source does not generate multiple images, the source position is rejected and re-sampled until either a set of two or more images are found, or the maximum number of iterations is reached and a new source redshift is drawn. Only images passing the selection criteria are used in modeling the synthetic clusters. Each set of images is indexed corresponding to their source, and the order of images are randomized before being passed to modeling to prevent potential bias. 

To test the reconstruction efficacy of {\tt HoliGRALE} as a function of the number of source systems, $N_{\rm S}$, 10 realizations of lensing systems were generated for each of the three clusters. The first realization contains 10 source systems, with each subsequent realization increasing by 10 systems, with a maximum of 100 source systems. None of the realizations share source systems. In total, each cluster contains 550 mock source systems. Each cluster is at a lens redshift of $z_{\rm d}=0.42$.

\subsection{Cluster 1: Harriet}\label{txt:c0}

This cluster was selected because of the elongated shape, but otherwise relatively smooth density map. It is Group $\#6$ in TNG-Cluster, and we used its X-Y projection. Harriet's mass within its virial radius, $R_{\rm 200}=943\,$ kpc $h^{-1}$, is $M_{\rm 200}=7.19\times10^{13}$ $M_\odot$ $h^{-1}$, and only particles within $R_{\rm 200}$ were used to construct the lens.

In calculating the surface mass density we included the contributions from all the components: dark matter, gas, stars, and black holes. To remove the discreteness effects of individual simulation particles the projected density was smoothed with a Gaussian filter of width $13.5\,$kpc. This scale was chosen to make the cluster isodensity contours appear smooth, in order to test {\tt HoliGRALE} in that regime.

{\tt HoliGRALE} uses galaxies in its reconstructions, and assigns mass to them based on their velocity dispersions. We identified cluster member galaxies as TNG subhalos that had at least $1.35\times10^{9}M_\odot$ of stellar mass within $r=10\,$kpc, and were within the strong lensing modeling window. There are a total of 8 of these. To convert stellar mass to the mock observed velocity dispersion we used a Faber-Jackson type relation, $\sigma=\sigma_0(L/L_0)^{3.4\,}$ km s$^{-1}$, where $L_0=6\times10^{10}L_\odot$, $\sigma_0=200$ km s$^{-1}$, and stellar mass-to-light ($M/L$) was assumed to be $0.7$.

To make the cluster more efficient as a lens we scaled up the surface mass density of all its mass components by a factor of 25. Since the only goal of these reconstructions is to test {\tt HoliGRALE} (and not to derive any astrophysical conclusions), rescaling the original TNG densities to match our purpose is acceptable.

Overall, Harriet is the simplest of the three clusters we used.

\subsection{Cluster 2: Maka Ska}\label{txt:c1}

The main driver in choosing our three clusters and setting their properties is to test {\tt HoliGRALE} in different strong lensing regimes. We therefore adjusted the properties of the clusters differently from each other. Here we describe only those properties that differ from Maka Ska compared to Harriet.

Maka Ska is Group $\#10$ in TNG-Cluster simulation. Its mass is
$M_{\rm 200}=3.37\times10^{13}$ $M_\odot$ $h^{-1}$ within the virial radius of $R_{\rm 200}=730\,$ kpc $h^{-1}$. As with Harriet, to reduce the effect of the particle discreteness we applied Gaussian smoothing to the 2D map of width $2.5\,$kpc. This small scale results in very wiggly isodensity contours, which we use to test {\tt HoliGRALE}'s ability to map out fine detail. We identified member galaxies as TNG subhalos with stellar mass $3\times10^9M_\odot$ within $10\,$kpc. Additionally, we imposed a compactness cut, selecting only those galaxies whose enclosed stellar mass within two different radii satisfied this criterion: $[M_\star(<50\,{\rm kpc})/M_\star(<10\,{\rm kpc})]<2$. The number of galaxies within the lens modeling window is 23. We used the same Faber-Jackson relation as for Harriet, but the stellar $M/L$ was assumed to be $0.4$.

The surface mass density of the cluster was scaled up by a factor of 42 to make the cluster a strong gravitational lens for sources as close as $z_{\rm s}\approx 0.7$.

\subsection{Cluster 3: Nokomis}\label{txt:c2}

This is Group $\#4$ in TNG-Cluster simulation. Its virial mass is 
$M_{\rm 200}=1.099\times10^{14}$ $M_\odot$ $h^{-1}$ within $R_{\rm 200}=1082\,$ kpc $h^{-1}$. Galaxies were selected as subhalos with stellar mass above $3\times10^8M_\odot$ within $10\,$kpc. As in Maka Ska we used a compactness criterion, selecting only those galaxies whose enclosed stellar mass satisfied $[M_\star(<50\,{\rm kpc})/M_\star(<10\,{\rm kpc})]<2.5$. Faber-Jackson relation is similar to the one in the previous clusters, but the stellar $M/L=0.4$, and $\sigma_0=180\,$ km s$^{-1}$. The total number of galaxies within the lens modeling window is 38.

In the case of Nokomis, to reduce the shot noise introduced by discrete particle numbers in simulations, we used a variable Gaussian smoothing length. For each kpc-sized square region we used a Gaussian width of $\sqrt{25^2/n}$, where $n$ is the total number of particles of all 4 types projected in that kpc$^2$. This allows regions of higher particle number density to have smaller smoothing length, and vice versa.

The final surface mass density of the cluster was scaled up by a factor of 12 to give it a reasonable strong lensing cross section for background sources.

\section{{\tt HoliGRALE} Methodology}\label{txt:holigrale}

{\tt HoliGRALE} is based on {\tt GRALE}, a publicly available free-form reconstruction code\footnote{\url{https://research.edm.uhasselt.be/~jori/grale2}}. The core components and methods in {\tt GRALE} are preserved, with the primary changes in {\tt HoliGRALE} being the treatment of parametric basis functions within the free-form skeleton. The diagnostic tests performed in this paper are designed to rigorously evaluate the new hybrid framework within the existing {\tt GRALE} architecture. We emphasize that {\tt HoliGRALE} is therefore an extension of {\tt GRALE}, and not an independent reconstruction code. 

In this section, we detail the sequential methodology of configuring a {\tt HoliGRALE} reconstruction. We first introduce the free-form {\tt GRALE} framework (Section \ref{txt:grale}), then describe the {\tt HoliGRALE} changes (Section \ref{txt:hybrid}), and finish with a description of fitness criteria and inputs (Section \ref{txt:fitness}).

\subsection{{\tt GRALE} Methods}\label{txt:grale}

{\tt GRALE} serves as the free-form foundation upon which {\tt HoliGRALE} is built. {\tt GRALE} uses an inversion technique based on a genetic algorithm that optimizes solutions to the lens equations. The only input used in {\tt GRALE} are the observed image positions of a lens, and no parametric information is inputted. Subsequently, the added features and diagnostic evaluations of {\tt GRALE} have been described in previous work \citep{liesenborgs06,liesenborgs07,zitrin10,mohammed14,meneghetti17,liesenborgs20,ghosh20}. Here, we will summarize the necessary features of {\tt GRALE} that are integral to {\tt HoliGRALE}.

A single {\tt GRALE} run begins first by defining an initially uniform grid populated by a basis set. Generally, projected Plummer spheres are used as the basis function, which have lens potentials of:
\begin{equation}\label{eq:plummerlpot}
    \psi(\boldsymbol{\theta}) = \frac{2GM D_\mathrm{ds}}{c^2 D_\mathrm{s} D_\mathrm{d}} \ln{(\boldsymbol{\theta}^2 + \theta_\mathrm{P}^2)},
\end{equation}
$\theta_\mathrm{P}$ is the characteristic angular width of the Plummer sphere, $M$ is its total mass, and $D$'s are angular diameter distances between the observer, {\it s}ource, and {\it d}eflector. Plummers are chosen due to their constant central density and smoothly and steeply decreasing density profile at large radii. Initially, the amplitudes of the Plummers are random.

The uniform grid of Plummers is in practice a dynamic grid that will evolve according to a genetic algorithm \citep{liesenborgs06}. The genetic algorithm seeks to optimize the weights of the basis functions. It does so by generating and selecting for realization solutions according to chosen fitness criteria (see Section \ref{txt:fitness}). Selected solutions, and crossover solutions combining trial solutions, become mutated and the entire process repeats until the overall fitness stops increasing substantially between generations \citep{liesenborgs06}. 

Once a peak fitness is reached, the dynamic grid is subdivided, with denser regions resolved even finer than less dense regions. On the new subdivided grid, Plummers are populated with proportional $\theta_{\rm P}$ to their grid cell size. A new best fitness solution is attained through the genetic algorithm. Subdividing the dynamic grid continues for a user-determined amount of subdivision steps. For this work, 10 subdivision steps are chosen with the grid increasing by $\sim$100 cells at each step. The output of a single {\tt GRALE} run is the subdivision step with the best (lowest) fitness criteria values. A typical {\tt GRALE} run generates a best fitting solution containing hundreds to thousands of Plummers from this subdivision process, with the smallest Plummer width of $\sim1''$. The final density distribution map can be computed from this dynamic grid of Plummers as a superposition.

Since each {\tt GRALE} run uses a different random realization of basis functions, the output of single runs will differ slightly. Therefore, the complete model is taken to be the average of an ensemble of {\tt GRALE} runs. It was shown that the mean model is an acceptable and often times better fitting model than any individual run \citep{liesenborgs06}. For this work, we follow recent {\tt GRALE} analyses and generate an ensemble of 40 {\tt GRALE} runs \citep{ghosh21,ghosh23,perera24,perera25a}, with the complete model being the average of these 40's lens potentials. We perform a source position optimization on this complete model to better fit observed image positions. This process is identical to the process detailed in \cite{perera24}.

\subsection{Adding Hybrid Components}\label{txt:hybrid}

{\tt HoliGRALE} builds upon the free-form {\tt GRALE} framework by introducing two primary changes: (1) introducing observationally motivated basis functions representing cluster member galaxies into the dynamic grid, and (2) explicitly subdividing the free-form components, and not any parametric components. We elaborate on these two improvements in this section.

The first change is the most important component of {\tt HoliGRALE}. Observationally motivated basis functions are now included in the initial dynamic grid. This functionally makes {\tt HoliGRALE} a hybrid methodology, as observational information about cluster member galaxies is now included into the genetic algorithm. These basis functions are optimized alongside the Plummers in the genetic algorithm. Explicitly adding non-Plummer components to the basis grid has been explored in previous studies \citep{liesenborgs20,perera24,liesenborgs24,perera25a}, however, these studies limited their addition to single or a few mass components. In this paper, we are testing the systematic inclusion of all observationally motivated mass components. Thus, we can no longer treat these basis function additions as small-scale perturbations to the free-form algorithm. The mass components we seek to include are all the cluster member galaxies observed in a given system.

Since the weights of the parametric basis functions are optimized, it is necessary to consider the subtle implications of incorporating them in a lens model. The main consideration is that since the weight of a basis function acts as a multiplicative factor, the optimization routine essentially solves for the amplitude of the basis function. In the case of Plummers (equation \ref{eq:plummerlpot}), weighting determines the total mass of the Plummer object. Weighting does not impact $\theta_{\rm P}$, which remains entirely determined by the size of the Plummer's subdivided grid cell. Therefore, the shape of the basis function density profile does not get rescaled. This is important when choosing profiles to represent cluster member galaxies. It is not as important to accurately measure the total mass of a cluster member galaxy, as it can be optimized in the reconstruction. It is very important, however, to approximate the shape of the galaxy density profile.

In previous work, various profiles have been used in hybrid {\tt GRALE} models. \cite{liesenborgs20} and \cite{liesenborgs24} represent a smaller scale substructure as non-singular isothermal ellipse (NSIE). \cite{perera24} models the central BCG as an elliptical Sersic lens. \cite{perera25a} represents galaxy scale components with Sersic and Navarro-Frenk-White (NFW) profiles. In these three cases, the shape properties of the respective profiles are either observed or estimated from scaling relations. For this paper, in order to evaluate {\tt HoliGRALE} without the bias of calibrations of these scaling relations or measurements, we represent galaxies as singular isothermal ellipses (SIE), with density profiles:
\begin{equation}
    \Sigma\left(\boldsymbol{\theta}\right) = \frac{\sigma_v^2 \sqrt{f}}{2GD_{\rm d}\sqrt{\theta_1^2 + f^2\theta_2^2}}\label{eq:SIS}
\end{equation}
The two parameters in the SIE profile are the velocity dispersion $\sigma_v$, which acts as the amplitude that will be optimized for in the genetic algorithm, and the ellipticity $f$ (minor-to-major axis ratio; $f=1$ corresponds to a circle). We initialize each galaxy in our models with an SIE based on the measured $\sigma_v$. If $\sigma_v$ is not measured, we scale it according to a Faber-Jackson like scaling relation, wherever applicable. If ellipticity information is available, then we include it. Otherwise, we take $f=1$, which reduces the SIE to the singular isothermal sphere (SIS). We emphasize that if the data is available, particularly slope or core radius information, then more realistic profiles can be included in {\tt HoliGRALE}.

The second major change in {\tt HoliGRALE} pertains to the subdivision of the dynamic grid. We include Plummers (representing the free-form components) and SIE (or SIS) lenses (representing the parametric components for galaxies) in the basis set. During the subdivision of the dynamic grid, we only subdivide the Plummer components. The reason for this is to ensure that each SIE is restricted to representing the galaxy mass information and that this stays roughly fixed during the optimization process. Subdividing the parametric components can lead to the SIE components over-representing the mass in their local regions. Early tests where parametric components are subdivided lead to a systematic inflation in their weights, leading to overly steep profiles for cluster members. Since the parametric components represent the observationally motivated light components of the lens (e.g., the baryons), this has the philosophical consequence of rendering the Plummers to be representative of the dark matter. In practice, this choice allows for continued flexibility in the model while anchoring the observed mass contribution in particular locations with SIEs.

\subsection{Inputs and Fitness Criteria}\label{txt:fitness}

The only inputs to {\tt HoliGRALE} are the multiple image positions, their respective source redshifts, and any parametric basis components as mentioned in the previous section. The genetic algorithm finds solutions by computing fitness criteria according to how well given solutions can reproduce these data \citep{liesenborgs07}. 



We use the default fitness criterion of {\tt GRALE} for these experiments. It should be noted that additional fitness criteria can be added at the user's discretion (see e.g., \citealt{liesenborgs07} and \citealt{liesenborgs09}). For this work, we focus mainly on the minimal example. We explore the time delay fitness criterion briefly below, and leave the additional cases for future studies.

In some cases, additional inputs can be added to the lens model. Time delay information is one particular example that has been studied before in {\tt GRALE} \citep{mohammed15,liesenborgs20,perera24}. If time delay information is available it can be used as a constraint in the genetic algorithm governed by the fitness function shown in equation 18 of \cite{liesenborgs20}. Notably, the time delay fitness is independent of the source position (assuming the model converges images well in the source plane), and this has been shown to be more accurate \citep{liesenborgs20}. When used, this fitness measure is generally prioritized the least. Since time delay information in clusters is very rare, has already been evaluated within {\tt GRALE} \citep{ghosh20}, and is not strongly prioritized, we do not include them as constraints in our analysis of synthetic clusters. We will explore its impact on a real lens system in Section \ref{txt:MACSJ0138}.

Lastly, we consider magnification in the lens modeling framework. Since the intrinsic brightness of sources is rarely constrained, it is generally not used as a constraint in lens modeling. Magnification ratios can be used, however, even if they are included, because of its nonlinear dependence on $\kappa$, including magnification as a constraint may lead to unphysical reconstructions. While we do not use magnification as a direct input constraint in {\tt HoliGRALE}, we seek to use magnification information (when available) as a form of model validation. It is known in cluster lens modeling that models exhibit a wide dispersion in magnification predictions even in low magnification ($\mu\lesssim2$) regions \citep{priewe17}. Therefore, it is important to evaluate a model's ability to accurately reconstruct magnification when available. Like with time delays, we explore this on a real lens system in Section \ref{txt:RXJ2129}.

\section{Metrics}\label{txt:metrics}

In this section, we define two classes of metrics that we will use to evaluate our lens models. The first are goodness of fit metrics, which explicitly try to quantify the accuracy of a lens model. The second are uncertainty metrics, which seek to quantify the error budget of a lens model. Given that gravitational lensing is a highly nonlinear problem subject to many degeneracies, these two classes are distinct. The ultimate goal of a lens model is to accurately reconstruct the image positions and the density distribution of the lens. These are somewhat independent tasks, subject to systematics. In addition, given our hybrid approach using an ensemble of lens models (see Section \ref{txt:holigrale}), measured quantities need to be evaluated across the entire model parameter space.

\subsection{Goodness of Fit Metrics}\label{txt:goodnessoffit}

It is necessary to define metrics that quantify how well a lens reconstruction fits to its input data. In the cluster lens modeling literature, the standard practice is to compute the lens plane root-mean-square (RMS):
\begin{equation}
    \Delta_{\rm RMS} = \sqrt{\frac{\sum_i^{N_{\rm im}} |\boldsymbol{\theta}_{i,\rm obs} - \boldsymbol{\theta}_{i,\rm rec}|^2}{N_{\rm im}}}\label{eq:rms}
\end{equation}
where $\boldsymbol{\theta}_{i,\rm obs}$ and $\boldsymbol{\theta}_{i,\rm rec}$  are the observed and reconstructed $i$th image positions, respectively. $N_{\rm im}$ is the total number of observed images used in the reconstruction.

An often overlooked limitation of $\Delta_{\rm RMS}$ as a metric is that it is uninformative to reconstructed image multiplicities. Sources where extra images (equivalently referred to as extraneous images in this work) are produced or where observed images are not reconstructed are inadequately handled in the definition. It is important to acknowledge instances of differing multiplicity on a case by case basis. Differences in image multiplicity between observations and reconstructions can be innocuous (e.g., extra images predicted near cluster centers that are unobserved due to the intracluster light) or potentially indicative of systematics (e.g., highly magnified extra images predicted close to cluster member density peaks, but not observed, hint at the under-constrained nature of the mass profile on small scales as this would imply a different mass profile shape). For these reasons, we explicitly include the number of extraneous reconstructed images and the number of images failed to be reproduced by our lens models in our results (see Table \ref{tab:summary}). It should be noted that the majority of these extraneous images are central images that are expected from gravitational lensing theory, but not observed due to their demagnification and location near the BCG. The value of $\Delta_{\rm RMS}$ in this paper, therefore, is only applicable to each model's $N_{\rm im}$ observed image positions.

While $\Delta_{\rm RMS}$ is a necessary metric in quantifying the accuracy of a lens model, we can go one step further in our analysis of synthetic clusters. Since we know the true $\kappa$ distribution, we quantify the accuracy of a given reconstruction according to the median percent error (MPE) between the model and true $\kappa$. This is an intuitive metric used to compare lens models (e.g., \citealt{bergamini23}). 

The percent error\footnote{As the form of PE is multiplied by 100, it is by definition a percentage. We quote it interchangeably in the text.} (PE) at a given location $\boldsymbol{\theta}$ between the convergence of the model ($\kappa_{\rm mod}$) and truth ($\kappa_{\rm true}$) is:
\begin{equation}
    {\rm PE}(\boldsymbol{\theta}) = \frac{|\kappa_{\rm mod}(\boldsymbol{\theta}) - \kappa_{\rm true}(\boldsymbol{\theta})|}{\kappa_{\rm true}(\boldsymbol{\theta})} \times 100 \label{eq:PD}
\end{equation}
The MPE is defined as the median of PE($\boldsymbol{\theta}$) at each grid point. In this way, the MPE is a pixel level comparison of how well the density is reconstructed by the lens model. Lower MPE values indicate better agreement with $\kappa_{\rm true}$.

\subsection{Uncertainty Metrics}\label{txt:uncertainty}

In general, $\Delta_{\rm RMS}$ acts as an estimate of the uncertainty of a lens model. This is the standard practice in cluster lens modeling. As discussed in Section \ref{txt:goodnessoffit}, there are important limitations to $\Delta_{\rm RMS}$, and it does not quantify how well $\kappa$ is reconstructed. Furthermore, previous diagnostic tests in lens modeling have found that a low $\Delta_{\rm RMS}$ is not necessarily indicative of a physical and accurate lens model \citep{ponente11,sendra14,williams18}. Therefore, $\Delta_{\rm RMS}$ alone is insufficient to quantify the uncertainty budget of a lens model. 

In parametric models, the typical way to quantify uncertainty, in addition to reporting $\Delta_{\rm RMS}$, is to provide the reduced $\chi^2$ or Bayesian Information Criterion (BIC). These allow for the consideration of the degrees of freedom in the model. These metrics are not applicable for free-form and some hybrid modeling techniques, including {\tt HoliGRALE}. In our case, we approach the problem of quantifying uncertainty in two ways: first by estimating the total error budget in the $\kappa$ reconstruction through the diagnostic experiments on synthetic clusters, and second, by estimating measured uncertainties on lensing observables by considering the lens model ensemble and its average.

First, as we will discuss in Section \ref{txt:synthetic}, we will determine the accuracy of {\tt HoliGRALE} based on the number of images used as constraints and the size of the cluster using the MPE. These tests on synthetic clusters can be extrapolated to real systems, upon which an order-of-magnitude estimate of the error budget in $\kappa$ can be attained. 

Second, when considering measured quantities of lensing observables (e.g., time delays, magnifications, etc.), treating the uncertainty on these predictions can be done in two ways. As mentioned in Section \ref{txt:holigrale}, {\tt HoliGRALE} creates an ensemble of lens models, where the full complete model is the average $\psi\left(\boldsymbol{\theta}\right)$ with optimized source positions. Because of this, there are in principle two ways to estimate uncertainty: (1) through the variance in the measured observable from the ensemble, or (2) through bootstrapping of the ensemble for the complete (averaged) model. The bootstrapping procedure involves randomly resampling the ensemble with replacement to compute the sampling distribution, the standard deviation of which acts as a measure of the uncertainty associated with the full complete model. Measuring uncertainty through the variance of the ensemble is best when we want to quantify how well {\tt HoliGRALE} is constraining a quantity. Bootstrapping the ensemble is best to quantify the uncertainty for the complete model. For our diagnostic experiments, (1) is most applicable since we directly want to know the systematics and error budget of {\tt HoliGRALE}. In practice, (2) is more practical as community users of {\tt HoliGRALE} models typically only work with the complete model. In this work, we will consider and present both of these values. In general, the precision and fidelity of a {\tt HoliGRALE} model can be evaluated based on the agreement between the ensemble-averaged and bootstrapped measurements.

\section{Reconstruction of Synthetic Clusters}\label{txt:synthetic}

\begin{table}
    \centering
	\caption{Summary of Synthetic Cluster Lens Reconstructions}
	\begin{tabular}{ccccccc} 
		\hline
		& $N_{\rm S}$ & $N_{\rm im}$ & $N_{\rm ex,im}$ & MPE & $\Delta_{\rm RMS}$ & Flags \\
		\hline
  		& 10 & 23 & 11 & 15.4 & $0.011''$ & 0\\
  		& 20 & 40 & 20 & 7.90 & $0.012''$ & 0 \\
  		& 30 & 71 & 29 & 6.70 & $0.056''$ & 0\\
  		& 40 & 92 & 40 & 5.90 & $0.052''$ & 0 \\
  		& 50 & 110 & 50 & 7.52 & $0.055''$ & 0 \\
  		Harriet & 60 & 148 & 62 & 6.09 & $0.059''$ & 0 \\
  	    & 70 & 173 & 73 & 5.47 & $0.145''$ & 0 \\
  		& 80 & 180 & 80 & 5.58 & $0.077''$ & 0 \\
  		& 90 & 213 & 91 & 5.37 & $0.071''$ & 0\\
  		& 100 & 250 & 100 & 4.66 & $0.087''$ & 0 \\
  		& 100 (FF) & 250 & 100 & 8.70 & $0.121''$ & 1\\
		\hline
  		& 10 & 27 & 10 & 26.2 & $0.034''$ & 0\\
  		& 20 & 51 & 24 & 25.5 & $0.317''$ & 1\\
  		& 30 & 76 & 32 & 18.8 & $1.720''$ & 0\\
  		& 40 & 97 & 50 & 17.0 & $0.270''$ & 0 \\
  		& 50 & 111 & 63 & 19.6 & $0.085"$ & 0 \\
  		Maka Ska & 60 & 137 & 81 & 19.9 & $0.245''$ & 1 \\
  	    & 70 & 165 & 75 & 17.3 & $0.242''$ & 1 \\
  		& 80 & 183 & 91 & 17.1 & $0.249''$ & 0 \\
  		& 90 & 233 & 97 & 16.6 & $0.542''$ & 0\\
  		& 100 & 222 & 115 & 15.7 & $0.227''$ & 1 \\
  		& 100 (FF) & 222 & 115 & 16.8 & $0.592''$ & 1\\
		\hline
  		& 10 & 36 & 14 & 39.1 & $2.19''$ & 0\\
  		& 20 & 61 & 37 & 31.3 & $0.180''$ & 0\\
  		& 30 & 78 & 38 & 13.6 & $0.268''$ & 0\\
  		& 40 & 126 & 53 & 14.5 & $0.224''$ & 1\\
  		& 50 & 150 & 63 & 11.4 & $0.170''$ & 1\\
  		Nokomis & 60 & 193 & 95 & 13.2 & $0.341''$ & 0\\
  	    & 70 & 192 & 94 & 11.0 & $0.566''$ & 0\\
  		& 80 & 254 & 126 & 14.9 & $0.363''$ & 0 \\
  		& 90 & 272 & 119 & 12.8 & $0.274''$ & 1\\
  		& 100 & 307 & 149 & 17.1 & $0.397''$ & 2\\
  		& 100 (FF) & 307 & 143 & 12.8 & $1.510''$ & 2\\
		\hline
	\end{tabular}\\
\medskip{A summary table of the lens modeling experiment carried out using synthetic clusters. The three sections are for the particular synthetic cluster being modeled (see Section \ref{txt:clusters}). Columns list the number of independent input source systems ($N_{\rm S}$), number of observed images used as input ($N_{\rm im}$), number of extraneous images (see Section \ref{txt:goodnessoffit}) predicted by the model ($N_{\rm ex,im}$), median percent error (MPE), $\Delta_{\rm RMS}$ (equation \ref{eq:rms}), and number of systems the model fails to reconstruct (``Flags''). The bottom row of each synthetic cluster shows, for comparison, the results for the free-form (FF) {\tt GRALE} model generated using the same input images as the $N_{\rm S}=100$ {\tt HoliGRALE} model.}
\label{tab:summary}
\end{table}

\begin{figure*}
\includegraphics[trim={1.5cm 0cm 2cm 0cm},clip,width=0.33\textwidth]{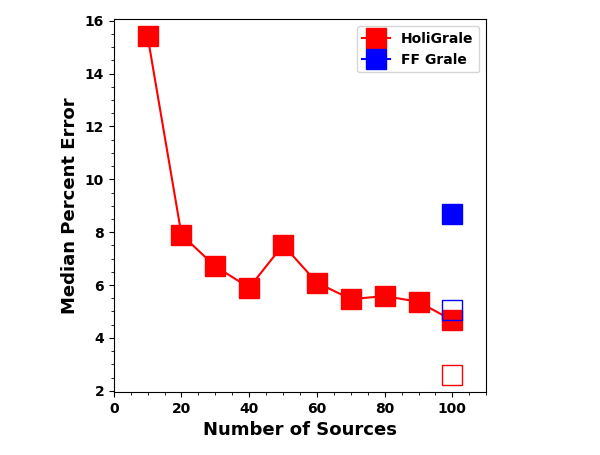}
\includegraphics[trim={1.5cm 0cm 2cm 0cm},clip,width=0.33\textwidth]{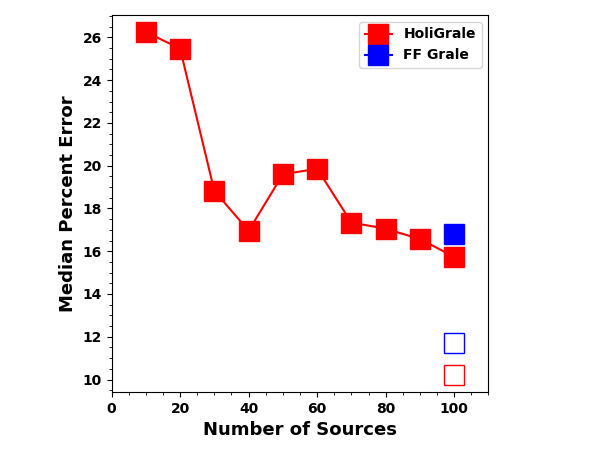}
\includegraphics[trim={1.5cm 0cm 2cm 0cm},clip,width=0.33\textwidth]{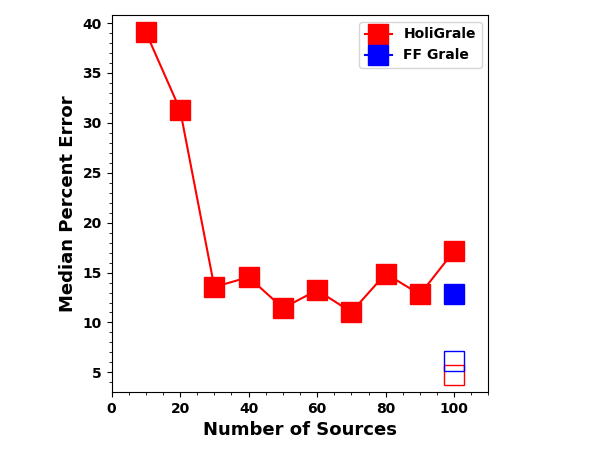}
\caption{The lens model MPE (see equation \ref{eq:PD} and Section \ref{txt:goodnessoffit}) plotted against the number of independent sources used as input for Harriet (\textit{Left}), Maka Ska (\textit{Middle}), and Nokomis (\textit{Right}). Closed squares indicate MPE calculated over the whole window of the cluster (see Figure \ref{fig:synthclusters}). Open squares are calculated for the inner region (defined by the images) as defined in Table \ref{tab:inner_mpd}.}
\label{fig:MPDs}
\end{figure*}

\begin{figure*}
\includegraphics[trim={0cm 0cm 0cm 0cm},clip,width=0.49\textwidth]{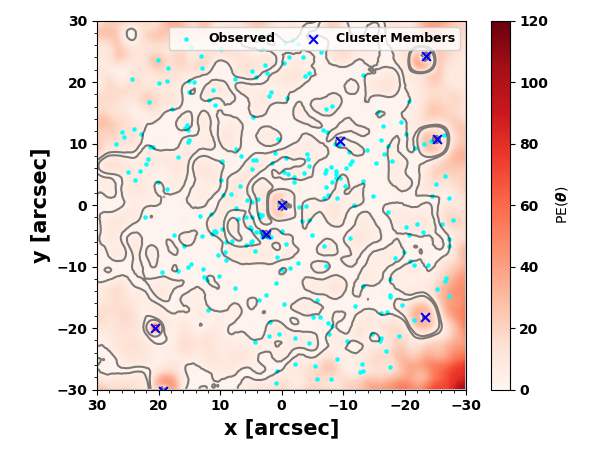}
\includegraphics[trim={0cm 0cm 0cm 0cm},clip,width=0.49\textwidth]{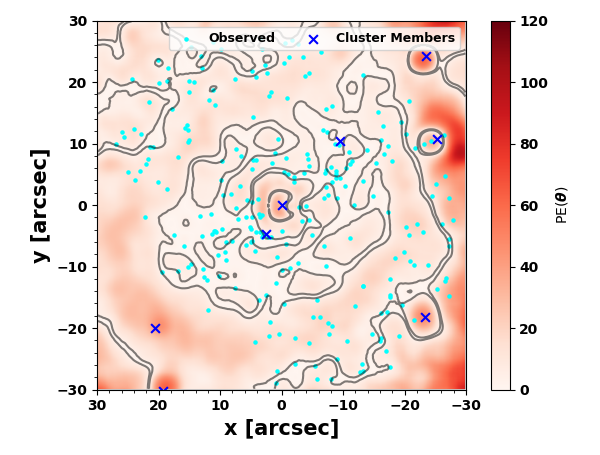}\\
\includegraphics[trim={0cm 0cm 0cm 0cm},clip,width=0.49\textwidth]{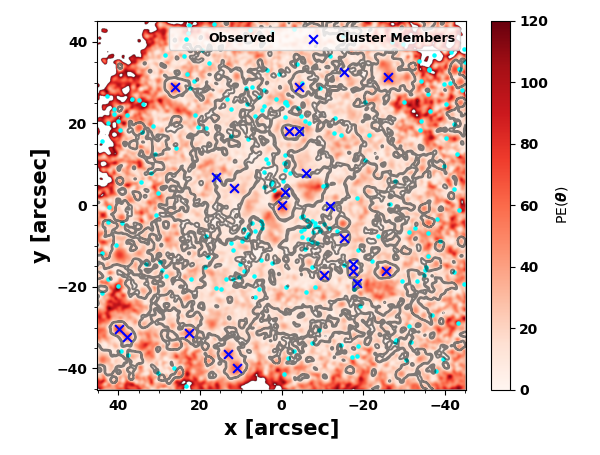}
\includegraphics[trim={0cm 0cm 0cm 0cm},clip,width=0.49\textwidth]{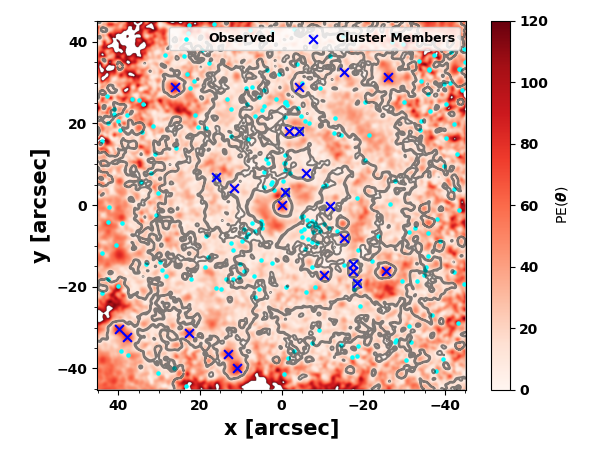}\\
\includegraphics[trim={0cm 0cm 0cm 0cm},clip,width=0.49\textwidth]{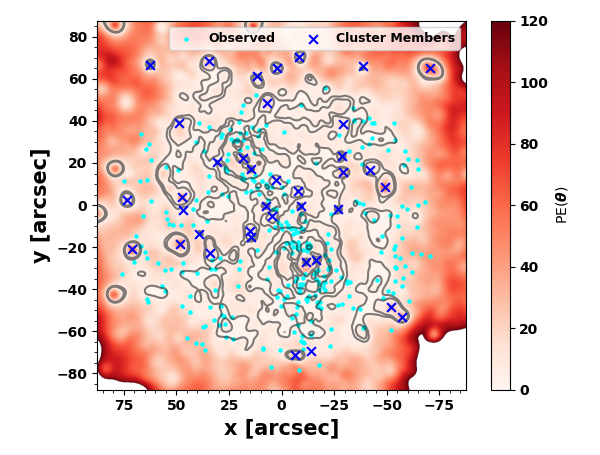}
\includegraphics[trim={0cm 0cm 0cm 0cm},clip,width=0.49\textwidth]{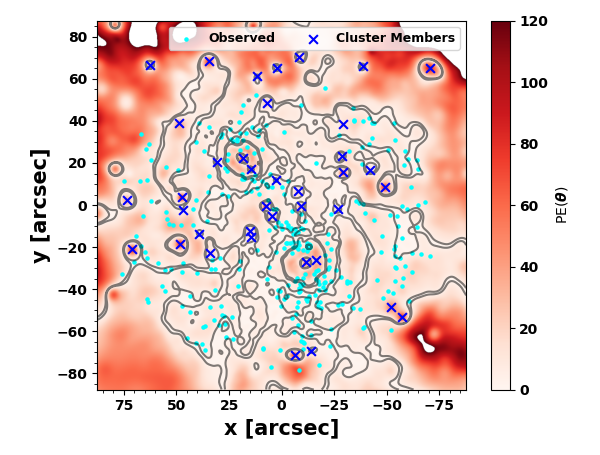}
\caption{Percent error (equation \ref{eq:PD}) maps for the $N_{\rm S} = 100$ {\tt HoliGRALE} (\textit{Left column}) and freeform {\tt GRALE} (\textit{Right column}) models of Harriet (\textit{Top row}), Maka Ska (\textit{Middle row}), and Nokomis (\textit{Bottom row}). The comparison is in the same window as those of Figure \ref{fig:synthclusters}. Darker shades of red indicate regions of greater disagreement between the model and true density distribution. Regions where PE$\left(\boldsymbol{\theta}\right)>120$ are masked out in order to better observe the fluctuations on small scales. Black contours indicate PE$\left(\boldsymbol{\theta}\right)=3$. Cyan dots and blue crosses denote image positions and cluster member galaxies, respectively.}
\label{fig:pdiffmaps}
\end{figure*}



To diagnostically evaluate {\tt HoliGRALE}, we perform the following experiment. For each of the three synthetic clusters described in Section \ref{txt:clusters}, we generate 10 {\tt HoliGRALE} models. The only difference between the 10 models is the image input: we steadily increase the number of background sources ($N_{\rm S}$) in increments of 10 (starting at 10 and finishing at 100). This is equivalent to increasing the number of input image positions steadily from the range of $\sim$20 to $\sim$300, depending on the size of the cluster. As a clarification, $N_{\rm S}$ corresponds to the number of independent background source systems. $N_{\rm S}$ is not necessarily indicating that all the systems have unique source redshifts. The source redshift distribution is normally distributed using the Box-Mueller algorithm with a mean of 2.8, standard deviation of 1, and upper limit of 6.5. All 10 {\tt HoliGRALE} models use the same cluster member galaxies as input for each synthetic cluster.

We seek to evaluate the accuracy of {\tt HoliGRALE} by making a plot of MPE with $N_{\rm S}$. In essence, this tracks how accurately {\tt HoliGRALE} reconstructs $\kappa$ as a function of the number of images used as input. To directly compare the performance of {\tt HoliGRALE} with free-form {\tt GRALE}, we also generate one {\tt GRALE} model using the exact same $N_{\rm S}=100$ sources. We are not comparing {\tt GRALE} across the whole range of $N_{\rm S}$, just at the expected most informative input, since the improved accuracy of {\tt GRALE} with image input was already established \citep{ghosh20}. We note that $N_{\rm S}=100$ is slightly larger than the $N_{\rm S}=79$ seen in MACSJ0416 \citep{rihtarsic25}, which is the current largest cluster lens in terms of observed images. Therefore, our experiment should serve as a useful diagnostic evaluation for future increased image data catalogs in the largest cluster lenses.

Table \ref{tab:summary} summarizes the results of all these lens models. We present our results and analysis of the accuracy of the mass and image reconstructions in Sections \ref{txt:massrecon} and Section \ref{txt:imagerecon}, respectively. Figure \ref{fig:MPDs} summarizes the quality of each reconstruction. Figure \ref{fig:pdiffmaps} shows the percent error maps for the $N_{\rm S}=100$ models.

\subsection{Mass Reconstruction}\label{txt:massrecon}

Figure \ref{fig:MPDs} depicts the MPE values for each model of each synthetic cluster as a function of $N_{\rm S}$. The MPE is calculated over the entire window of the cluster (solid points of Figure \ref{fig:MPDs}), which is the same view as shown in Figure \ref{fig:synthclusters}. In each cluster, the MPE remains roughly stable once $N_{\rm S}$ increases beyond 20–30. For all subsequent models, the MPE remains roughly stable over the full range of $N_{\rm S}$. We define this stable region as the plateau, which occurs for models with $N_{\rm S}\geq20$ for Harriet, and $N_{\rm S}\geq30$ for Maka Ska and Nokomis. The full percent error maps are presented for select models in Figure \ref{fig:pdiffmaps}, and the reconstructed density distributions are shown in Appendix \ref{txt:pdiffmaps}.

This plateau in MPE allows us to quantify the expected error budget in the $\kappa$ reconstruction of {\tt HoliGRALE} for each type of cluster. We define this error budget to be the mean of the MPE plateau, with uncertainty the standard deviation. For Harriet-like clusters (unimodal with few cluster members), {\tt HoliGRALE} is able to reconstruct the entire density distribution to $6.1\pm1.0\%$ accuracy. For Maka Ska-like clusters (unimodal with many cluster members), the error budget is $17.6\pm1.4\%$. For Nokomis-like clusters (bimodal with many cluster members), the error budget is $13.6\pm1.8\%$. Importantly, in all three cases, {\tt HoliGRALE} exhibits an improvement over or equivalence with {\tt GRALE} in the reconstruction of the mass distribution. We note that Maka Ska has a more rough density distribution, which contributes to its somewhat larger MPE result.

If we bound the PE comparison region to the inner region, we can probe the goodness of the reconstruction in the densest cluster regions, close to the center and where the image positions are more concentrated. For each, we restrict the inner comparison region to be half of the window described in Appendix \ref{txt:pdiffmaps} (e.g. for Harriet the window is reduced from the depicted 40x40 square arcsec to 20x20 square arcsec, with the same center). These results are summarized in Table \ref{tab:inner_mpd}, and shown as open points in Figure \ref{fig:MPDs}. For Harriet, Maka Ska, and Nokomis, the MPE error budget reduces to $3.66\pm0.68\%$, $12.0\pm1.0\%$, and $5.03\pm0.70\%$, respectively. The uncertainties again correspond to the standard deviation of the MPE plateau. We highlight that for the $N_{\rm S}=100$ model MPE is 2.60\%, 10.2\%, and 4.76\%, for Harriet, Maka Ska, and Nokomis, respectively. The freeform {\tt GRALE} error budget becomes 5.05\%, 11.7\%, and 6.12\% for Harriet, Maka Ska, and Nokomis, respectively. 

From this MPE analysis, we can conclude that for simple, few component (Harriet-like) clusters, {\tt HoliGRALE} represents an improvement over {\tt GRALE}. For more complex clusters, like those similar to Maka Ska and Nokomis, {\tt HoliGRALE} reconstructs the mass equally as well as {\tt GRALE}. In the inner regions of clusters, where images are more prevalent and where the structure is somewhat more complex, {\tt HoliGRALE} is an improvement over {\tt GRALE} across the board. We expect the error budget in the mass reconstructions to be $\lesssim10\%$. 

It is worth investigating the impact of cluster members on {\tt HoliGRALE} with these conclusions in mind. As the addition of cluster members as SIS is the primary difference of {\tt HoliGRALE} with freeform {\tt GRALE}, we can directly quantify the improvement of the model in regions near the cluster members. For the $N_{\rm S}=100$ models of {\tt HoliGRALE} and {\tt GRALE}, we compute the MPE in regions $<5''$ from each cluster member. For Harriet, Maka Ska, and Nokomis, {\tt HoliGRALE} ({\tt GRALE}) finds a MPE of 6.08 (12.6), 14.1 (15.8),  and 7.96 (10.8), respectively. In each synthetic cluster, {\tt HoliGRALE} is reconstructing the mass near cluster members better than {\tt GRALE}. This result is perhaps to be expected, however, it serves as confirmation that the addition of cluster members is a driving factor in the improvement of {\tt HoliGRALE} over {\tt GRALE}. 

\begin{table}
\centering
\caption{MPE comparison in the inner reconstruction regions.}
\label{tab:inner_mpd}
\setlength{\tabcolsep}{2.45pt}
\begin{tabular}{@{}lccc@{}}
\hline
 & Harriet & Maka Ska & Nokomis \\
\hline
Inner region & $20''\times20''$ & $22.6''\times22.6''$ & $50''\times50''$ \\
MPE error budget (\%) & $3.66\pm0.68$ & $12.0\pm1.0$ & $5.03\pm0.70$ \\
{\tt HoliGRALE} $N_{\rm S}=100$ MPE & $2.60$ & $10.2$ & $4.76$ \\
FF {\tt GRALE} $N_{\rm S}=100$ MPE & $5.05$ & $11.7$ & $6.12$ \\
\hline
\end{tabular}\\
\medskip{MPE summary of the model reconstructions in the inner regions of each synthetic cluster. The MPE error budget is defined by the same $N_{\rm S}$ models in Figure \ref{fig:MPDs} that make up the plateau (see text).}
\end{table}

\subsection{Image Reconstruction}\label{txt:imagerecon}

We now seek to evaluate the accuracy of the image position reconstructions. This is a related but independent goal of lens modeling, as it is well established that the model $\Delta_{\rm RMS}$ is expected to rise with $N_{\rm im}$, since it is easier to fit fewer image positions due to the wider range of approximate degeneracies \citep{johnson16,vegaferrero19,ghosh20}. As discussed in Section \ref{txt:uncertainty}, $\Delta_{\rm RMS}$ (equation \ref{eq:rms}) alone is insufficient to evaluate a lens model. Therefore, we will scrutinize image reconstructions beyond just the typical $\Delta_{\rm RMS}$.

Looking at the $\Delta_{\rm RMS}$ results in Table \ref{tab:summary} we can observe that generally $\Delta_{\rm RMS}$ does increase with $N_{\rm im}$ as expected. In all three synthetic clusters, the $N_{\rm S}=100$ {\tt HoliGRALE} models all exhibit a lower $\Delta_{\rm RMS}$ than their freeform {\tt GRALE} counterparts. Therefore, from a first glance view, {\tt HoliGRALE} appears to do better than {\tt GRALE}. In the following sections, we scrutinize the failure modes of these lens models (Section \ref{txt:failuremodes}) and image reconstruction correlations (Section \ref{txt:correlations}).

\subsubsection{Failure Modes}\label{txt:failuremodes}

Another important evaluation metric is to understand where a lens model fails to reconstruct an image system. We have quantified this as the ``Flags'' parameter in Table \ref{tab:summary}. It turns out that failing to reconstruct an image system properly is very rare, with no model regardless of $N_{\rm S}$ consisting of greater than 2 flags. Investigating these flags more closely lead to valuable insights into the shortcomings of our lens models.

Starting with Harriet, only the $N_{\rm S}=100$ freeform {\tt GRALE} model contains a failed image reconstruction. Notably, the {\tt HoliGRALE} counterpart does not contain this flag. The culprit for this case is one of the sources coincidentally forms a quad lens system centered on the galaxy at (-25,10) (see Appendix \ref{txt:pdiffmaps}). The explicit inclusion of this cluster member galaxy in {\tt HoliGRALE} is why it gets properly reconstructed. This indicates that small scale mass features can be reconstructed in {\tt HoliGRALE}.

In Maka Ska, the true mass distribution is much less smooth than the other two. This causes a much more complex small scale structure, with a very windy critical curve. In rare cases, images that form close to this critical curve are improperly reconstructed, due to the images being split by the winding critical curve. Therefore, the inability of {\tt HoliGRALE} to properly reconstruct these systems is due to the underconstrained nature of these regions, stemming from the much more rough density distribution.

As an example of this, model $N_{\rm S} = 30$ finds a large $\Delta_{\rm RMS}$. This is due to one source that contains one of these counterimage pairs produced from the critical curve winding effect. The model predicts an image splitting for the wrong set of images (i.e., a set of images is split, but the model splits a different counterimages, preserving multiplicity but inflating $\Delta_{\rm RMS}$), hence not constituting a failed reconstruction, but resulting in a very large RMS. Masking this source out yields $\Delta_{\rm RMS} = 0.511''$, a much more reasonable result. Since this is the only case in the entire suite of models for Maka Ska, we deem it an outlier. In general, all flags in the Maka Ska suite of reconstructions are the result of image splitting by the winding critical curve. Since this is a rare phenomenon that is highly underconstrained, we do not draw a conclusion based solely off of it.

In Nokomis, similar scenarios to those discussed in the previous paragraphs are observed. Two large $\Delta_{\rm RMS}$ reconstructions are recovered due to image splitting for the wrong set of images. When masking this source out, $\Delta_{\rm RMS}$ reduces to $0.141''$ and $0.536''$ for Nokomis $N_{\rm S}=10$ and $N_{\rm S}=100$ {\tt GRALE}, respectively. The flags for Nokomis are entirely caused by a small mass clump present near the massive cluster member at ($-12'', -27''$). This small mass clump lies below the compactness criterion for this cluster (see Section \ref{txt:c2}), and so was not included in the list of cluster members. Despite this, it contains enough mass to alter the deflection field such that sources near it can be split radially. Since it is not included in the reconstructions, its existence cannot be recovered by either method. We deem this feature an artifact stemming from our compactness criterion to define cluster galaxies. Given its small mass, we do not expect it to impart any significant bias in any of our other results.

From this deeper scrutiny of the failure modes of {\tt HoliGRALE}, we conclude that signatures of small scale structure that are present in the images used in the reconstruction are of great importance in {\tt HoliGRALE}. This is especially true for galaxy-scale lensing features within the cluster, akin to the one seen in Harriet. This is consistent with the results from Section \ref{txt:massrecon}. {\tt HoliGRALE} (and {\tt GRALE}) fails primarily when images are split nontrivially in regions away from cluster members. 

\subsubsection{Correlations}\label{txt:correlations}

\begin{figure}
\includegraphics[trim={0cm 0cm 0cm 0cm},clip,width=0.49\textwidth]{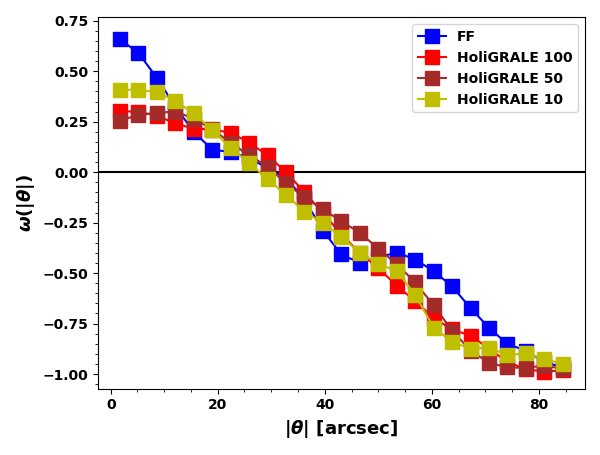}
\includegraphics[trim={0cm 0cm 0cm 0cm},clip,width=0.49\textwidth]{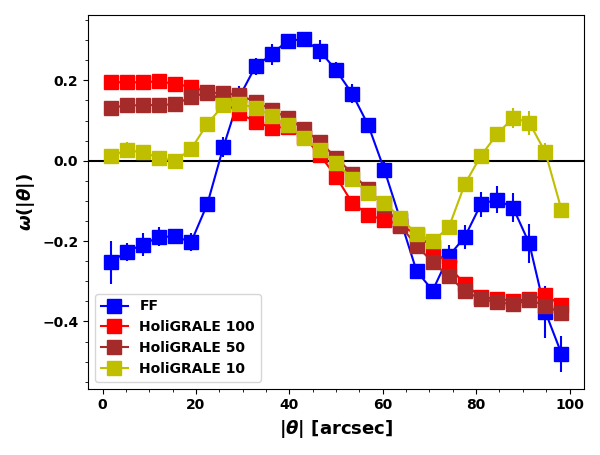}
\includegraphics[trim={0cm 0cm 0cm 0cm},clip,width=0.49\textwidth]{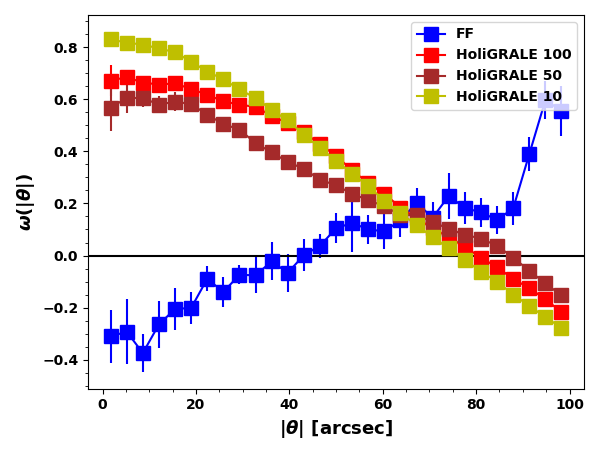}
\caption{Cross-correlation functions $\omega\left(|\boldsymbol{\theta}|\right)$ (equation \ref{eq:correlationfunc}) for Harriet (\textit{Top}), Maka Ska (\textit{Middle}), and Nokomis (\textit{Bottom}). Freeform (FF) {\tt GRALE} $N_{\rm S}=100$, {\tt HoliGrale} $N_{\rm S}=100$, {\tt HoliGrale} $N_{\rm S}=50$, and {\tt HoliGrale} $N_{\rm S}=10$ are shown in blue, red, brown, and yellow, respectively. Error bars indicate the standard deviation of different normalizations of $\omega\left(|\boldsymbol{\theta}|\right)$. }
\label{fig:crosscorr}
\end{figure}

Now, we ask the question specifically of how important the image position input to our models affects the quality of the reconstruction. To answer this question, we compute the cross-correlation function between the input image number density $n\left(\boldsymbol{\theta}\right)$ and percent error PE$(\boldsymbol{\theta})$:
\begin{equation}
    \omega\left(|\boldsymbol{\theta}|\right) = \frac{\sum_{(i,j)\in|\boldsymbol{\theta}|}n\left(\boldsymbol{\theta}_i\right)/{\rm PE}(\boldsymbol{\theta}_j)}{\sum_{(i,j)\in|\boldsymbol{\theta}|}n\left(\boldsymbol{\theta}_{r}\right)/{\rm PE}(\boldsymbol{\theta}_j)} - 1\label{eq:correlationfunc}
\end{equation}
where $|\boldsymbol{\theta}|=|\boldsymbol{\theta}_i-\boldsymbol{\theta}_j|$. To evaluate equation \ref{eq:correlationfunc}, we first calculate $n\left(\boldsymbol{\theta}\right)$ on the same grid as PE$(\boldsymbol{\theta})$. The grid is equivalent to those of Figure \ref{fig:synthclusters}, but with a smaller resolution of 100$\times$100 pixels to speed up calculations. $n\left(\boldsymbol{\theta}\right)$ is also smoothed using Gaussian kernel density estimation. For each bin $|\boldsymbol{\theta}|$ we compute equation \ref{eq:correlationfunc}, where we normalize against a random image plane position $\boldsymbol{\theta}_{r}$. If $\omega\left(|\boldsymbol{\theta}|\right)>0$, it can be interpreted that the image number density is anti-correlated with PE$(\boldsymbol{\theta})$, indicating that a higher density of input images is more likely to be associated with a more accurate reconstruction on scales of $|\boldsymbol{\theta}|$. The inverse is true when $\omega\left(|\boldsymbol{\theta}|\right)<0$.

Figure \ref{fig:crosscorr} shows the cross-correlation functions for the three synthetic clusters. For Harriet, the relations are similar across different models, all consistent with more accurate reconstructions in regions of higher image number density. For Maka Ska and Nokomis, the freeform {\tt GRALE} lens models' reconstruction accuracies show little to no dependence on the image number density. However, for all the {\tt HoliGRALE} models shown, there is a strong relation between image number density and reconstruction accuracy.

These results allow us to interpret the results of {\tt HoliGRALE} models in a specific way. For a given {\tt HoliGRALE} reconstruction, we can assume that the most accurately reconstructed regions of the cluster will be in the regions of higher image number density. From Figure \ref{fig:crosscorr}, this appears to be true regardless of the complexity of the cluster. Since {\tt GRALE} does not exhibit this universally, we can also conclude that {\tt HoliGRALE} will give more reliable constraints on the mass distribution near cluster members.

It should be noted that $\omega\left(|\boldsymbol{\theta}|\right)$ is dependent on the resolution that PE$(\boldsymbol{\theta})$ is computed on. We explore this directly in Appendix \ref{txt:crosscorr50}, and our results support our aforementioned conclusions. If the resolution used to construct the correlations is made coarser, the shape of $\omega\left(|\boldsymbol{\theta}|\right)$ for {\tt GRALE} becomes very similar to that of {\tt HoliGRALE}.

\subsection{Magnification and Time Delay Reconstruction}\label{txt:mutdrecon}

\begin{figure*}
\includegraphics[trim={0cm 0.5cm 0cm 1.9cm},clip,width=0.33\textwidth]{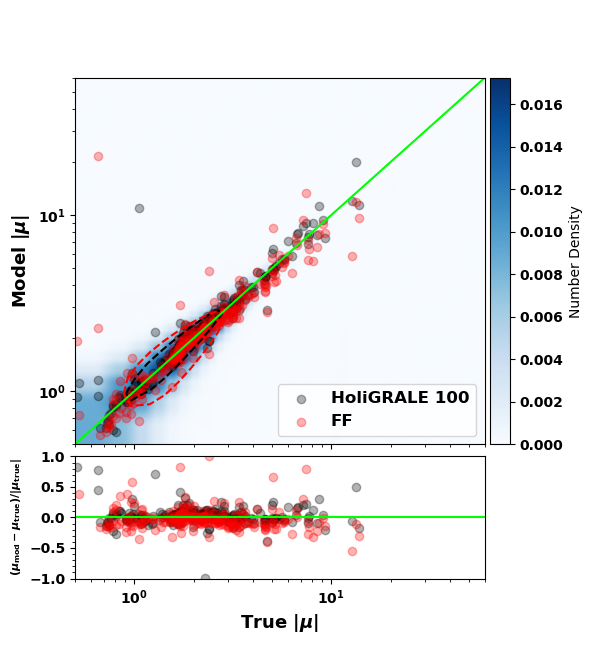}
\includegraphics[trim={0cm 0.5cm 0cm 1.9cm},clip,width=0.33\textwidth]{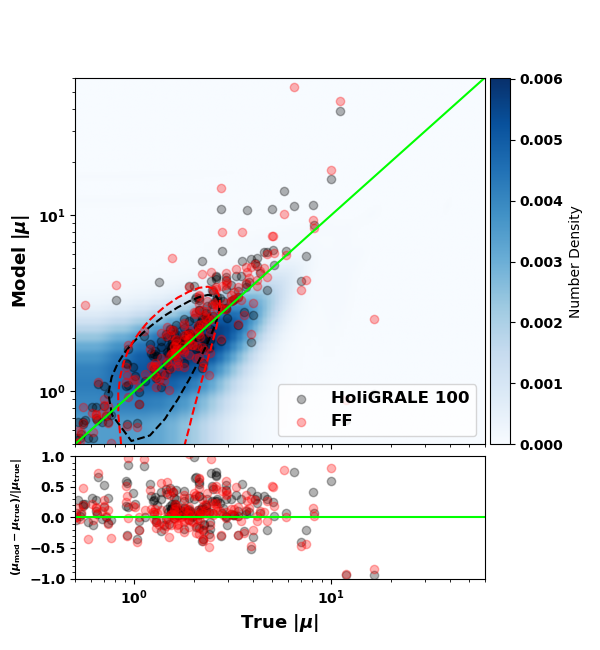}
\includegraphics[trim={0cm 0.5cm 0cm 1.9cm},clip,width=0.33\textwidth]{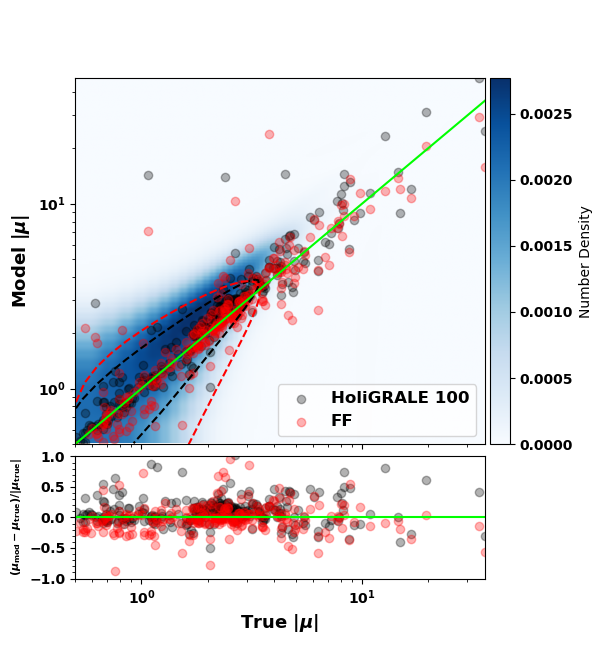}
\caption{Model reconstructed absolute magnification versus true absolute magnification for Harriet (\textit{Left}), Maka Ska (\textit{Middle}), and Nokomis (\textit{Right}). Fractional uncertainties are also shown beneath each plot. The green line indicates unity. Black and red points indicate individual image magnifications for {\tt HoliGRALE} $N_{\rm S}=100$ and freeform {\tt GRALE}, respectively. The blue colormap shows the number density of images to guide the eye as to where magnifications of images are generally concentrated. Contours indicate the middle 30\% confidence level for both models. As can be seen, this interval is smaller for {\tt HoliGRALE} in each cluster.}
\label{fig:muresid}
\end{figure*}

\begin{figure*}
\includegraphics[trim={0cm 0.5cm 0cm 1.9cm},clip,width=0.33\textwidth]{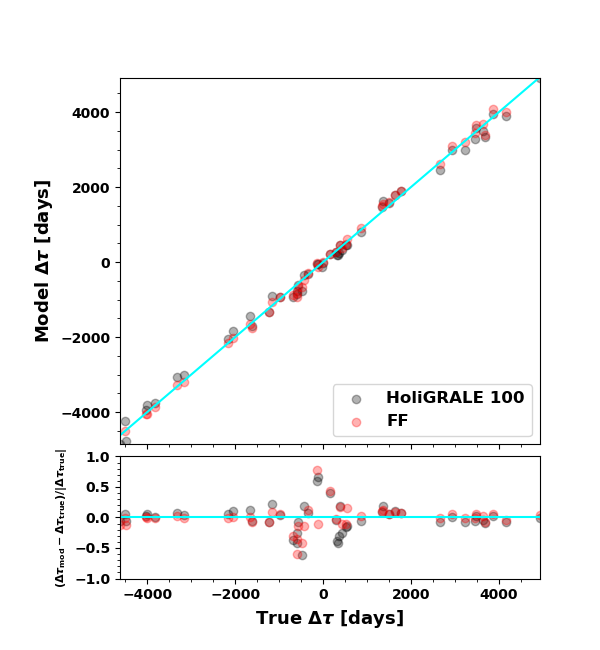}
\includegraphics[trim={0cm 0.5cm 0cm 1.9cm},clip,width=0.33\textwidth]{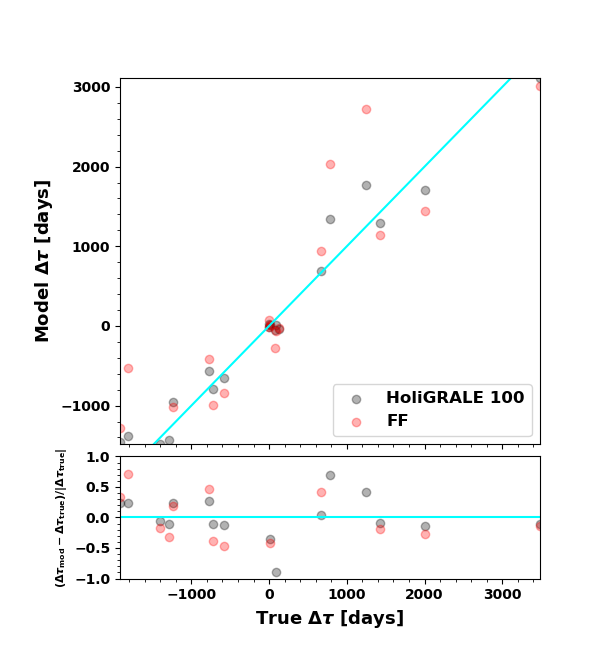}
\includegraphics[trim={0cm 0.5cm 0cm 1.9cm},clip,width=0.33\textwidth]{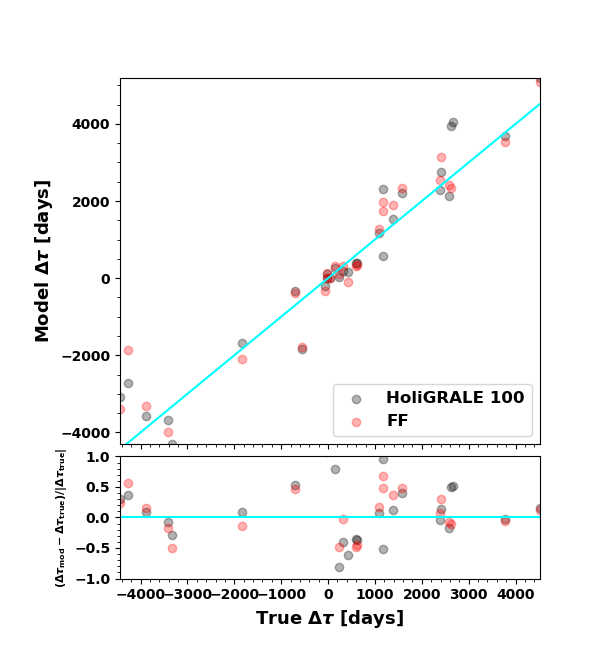}
\caption{Similar to Figure \ref{fig:muresid}, but for time delays under 5000 days. Model versus true time delays for Harriet (\textit{Left}), Maka Ska (\textit{Middle}), and Nokomis (\textit{Right}), with fractional uncertainties underneath each plot. Red and black indicate individual time delays for {\tt HoliGRALE} $N_{\rm S}=100$ and freeform {\tt GRALE}, respectively. The cyan line indicates unity. No number density is shown in this case due to the small number of applicable points.}
\label{fig:tdresid}
\end{figure*}

Lastly, we evaluate how well {\tt HoliGRALE} is able to reconstruct magnifications and time delays. While observations of these quantities are rare, quantifying the efficacy of {\tt HoliGRALE} in their reconstruction is still a worthwhile exercise to understand the systematics of the lens modeling algorithm \citep{meneghetti17,cha26}.

For magnifications and time delays, we compute the true values ($\mu_{\rm true}$ and $\Delta\tau_{\rm true}$) at each observed image position. For $\Delta\tau_{\rm true}$, time delays are computed relative to the first listed image in the catalog for each image system. As this is a blind analysis, this is effectively random, mimicking real observations of time delays where generally the first discovered time delay is never guaranteed to be the physically first arriving. 

For each lens model, we compute the reconstructed magnifications and time delays ($\mu_{\rm mod}$ and $\Delta\tau_{\rm mod}$) for each reconstructed image position\footnote{This is the prescription detailed in equation 11 of \cite{ghosh20}, regarding $p^{RT}$.}. We do not consider failed reconstructions for this exercise. To evaluate the goodness in reconstructing the magnifications and time delays, we compute the root-mean-square deviation (RMSD):
\begin{equation}
    {\rm RMSD}(x) = \sqrt{\frac{1}{N}\sum\left(\frac{x_{\rm mod} - x_{\rm true}}{x_{\rm true}}\right)^2}\label{eq:rmsd}
\end{equation}
where $N$ is the number of observed $x$. $x$ signifies either the magnification or the time delay. The analysis of time delay reconstructions has been previously analyzed for freeform {\tt GRALE} \citep{ghosh20}, but not for other lens modeling algorithms \citep[e.g.,][]{meneghetti17,cha23}. Our total results are summarized in Table \ref{tab:mutdresid} in Appendix \ref{txt:mutdtab}.

For magnifications, our results are shown in Figure \ref{fig:muresid}, where we present the model versus true $\mu$, along with the fractional uncertainties around unity. As can be seen, in all three synthetic clusters, the magnifications of images are accurately reconstructed. Furthermore, for the $N_{\rm S}=100$ reconstructions, RMSD($\mu$) is lower for {\tt HoliGRALE} in all three clusters (see Table \ref{tab:mutdresid}), indicating that {\tt HoliGRALE} is generally more accurate than freeform {\tt GRALE}. 

Since $\mu$ is not uniformly distributed over the entire domain, we also consider the RMSD of the subset of images that contain $|\mu|>5$. These highly magnified images are closer to the critical curve, and generally are the ones to be observed in practice. As seen in Table \ref{tab:mutdresid}, generally RMSD ($|\mu|>5$) is smaller than the entire set of images. This indicates that both {\tt HoliGRALE} and {\tt GRALE} are better at constraining highly magnified images than images with lower magnification. This is not surprising, as highly magnified images are more likely to exist in the inner regions of the cluster, where there is a high probability of a large image number density. We note, however, that the uncertainties on $\mu$ for these highly magnified images can be very large due to their positions close to critical curves. This is explored more in Section \ref{txt:RXJ2129uncertainties}.

We find that for most models the time delays are very well reconstructed (see Table \ref{tab:mutdresid}). Like with magnifications, {\tt HoliGRALE} is more accurate than {\tt GRALE}. We check the subset for observable time delays in this case, defined as images with $\Delta\tau<5000$\,days. This limit is motivated by observations; the longest observed time delay is $\sim2500$\,days \citep{munoz22}, and it is expected in the near future that time delays of $\sim4000$\,days will be observed (see e.g., \citealt{suyu26} and Section \ref{txt:tdpred}). We present these results in Figure \ref{fig:tdresid}. The RMSD for this subset is larger than the whole sample, but not by a significant amount in most cases.

In Harriet, time delays are very well reconstructed by both {\tt HoliGRALE} and {\tt GRALE}. In Maka Ska and Nokomis, the time delays are more mixed. Generally the RMSD is larger in Maka Ska and Nokomis than in Harriet, with {\tt HoliGRALE} outperforming {\tt GRALE}. Despite this, Figure \ref{fig:tdresid} still shows the images with $\lvert\Delta\tau\rvert<5000$\,days lie roughly along the line of unity, indicating that the lens models are recovering the correct trend. We conclude from these tests that for Harriet-like clusters, {\tt HoliGRALE} and {\tt GRALE} are able to accurately reconstruct time delays. For more complex, multi-component clusters, time delays are constrained more poorly, with {\tt HoliGRALE} doing slightly better than {\tt GRALE}. We note that these results are for lens models that do not include time delay information as a constraint.

From these results, we now apply {\tt HoliGRALE} to three real cluster lenses with recent high quality observations.

\section{Reconstructions of Real Cluster Lenses}\label{txt:real}

With {\tt HoliGRALE} established as an improvement over {\tt GRALE} across multiple realistic mock lenses of varying complexity, we now apply {\tt HoliGRALE} to three real cluster lenses. These systems were chosen because each contains a unique property that we wish to evaluate {\tt HoliGRALE} on. RX\,J2129.7$+$0005 (hereafter RXJ2129) contains SN\,2022riv which has a well constrained magnification \citep{dhanasingham26}. MACS\,J0138.0$-$2155 contains two supernovae with measured time delays \citep{rodney21,pierel24}. AS1063 is a massive HFF lens that contains a large central core with potential implications for dark matter \citep{limousin22,diego26b}. We analyze all three in greater detail in the following sections. For these three lenses, we assume a cosmology of $H_0=70$ km s$^{-1}$ Mpc$^{-1}$, $\Omega_{\rm m,0}=0.3$, and $\Omega_\Lambda=0.7$ to ensure consistency with the literature.

\subsection{RX\,J2129.7$+$0005}\label{txt:RXJ2129}

\begin{figure*}
\includegraphics[trim={0cm 0cm 0cm 0cm},clip,width=0.49\textwidth]{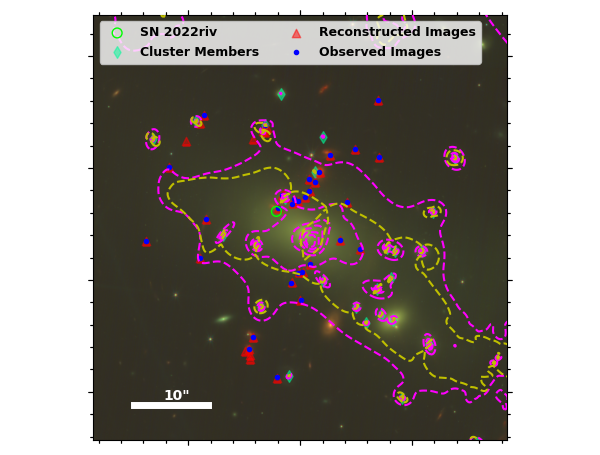}
\includegraphics[trim={0cm 0cm 0cm 0cm},clip,width=0.49\textwidth]{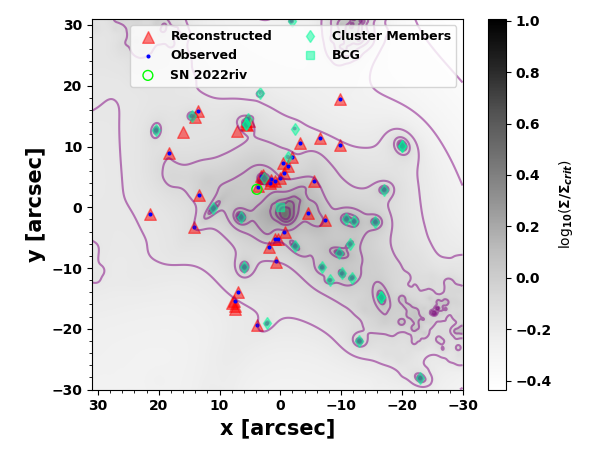}
\caption{{\tt HoliGRALE} lens modeling of RXJ2129. \textit{Left:} JWST RGB color image of RXJ2129. Overplotted magenta and yellow contours indicate the model density distribution and critical curve for $z=1.522$, respectively. \textit{Right:} Full density distribution of the lens model, with purple density contours separated by $\Delta\log_{10}\kappa=0.1$. In both panels, SN\,2022riv, the object of interest, is denoted by a bright green circle. Cluster members, reconstructed images, and observed images are shown as green diamonds, red triangles, and blue dots, respectively. The BCG is shown at the center as a green square.}
\label{fig:RXJ2129}
\end{figure*}

\subsubsection{Lens Modeling Summary}

RXJ2129 ($z_{\rm d}=0.234$) has been previously modeled using Hubble Space Telescope (HST) data \citep{richard10,zitrin15,desprez18,caminha19,okabe20,jauzac21}. James Webb Space Telescope (JWST) data enabled the discovery of a compact high redshift ($z=9.51$) lensed galaxy system \citep{williams23}, three new image systems within a highly magnified lensed arc, and a multiple imaged Type Ia supernova, SN\,2022riv \citep{dhanasingham26}. As a result, there is renewed interest in this system. In this section, we analyze the {\tt HoliGRALE} lens model of RXJ2129, which was used in the analysis presented in \cite{dhanasingham26} to constrain the magnification ($\mu$) of SN\,2022riv. Therefore, we use SN\,2022riv as a case study in understanding how well {\tt HoliGRALE} can reconstruct $\mu$.

Our {\tt HoliGRALE} model utilizes 31 spectroscopically confirmed image positions from 9 independent source planes as input \citep[see Table 13 of][]{dhanasingham26}. We also include 68 cluster member galaxies \citep{jauzac21}, which are included in the basis set as SIEs since they contain measurements of ellipticity. We direct the reader to Section 4.1.3 of \cite{dhanasingham26} for more specific details of the lens model. Our model achieves $\Delta_{\rm RMS}=0.09''$. Figure \ref{fig:RXJ2129} shows the density distribution for this model.

The analysis presented in \cite{dhanasingham26} measures $\mu = -5.35\pm1.01$ for SN 2022riv. Our model finds $\mu=-15.39\pm0.85$, in $7.6\sigma$ tension with the observed value. Accounting for microlensing effects, however, reduced this tension to $<1\sigma$. The large disagreement is somewhat surprising considering our expected error budget as we examined in Section \ref{txt:synthetic}. In the context of our synthetic clusters, RXJ2129 is closest to a Maka Ska-like cluster, given its unimodality and large number of cluster members. From this order of magnitude comparison, we should expect $\sim20\%$ disagreement between our model and the true mass distribution. However, the image number density near SN\,2022riv is very high; around a third of the observed images lie within $5''$ of SN 2022riv. From our results in Section \ref{txt:correlations}, we should therefore expect higher accuracy in the reconstruction in this region, which is not supported by our magnification reconstruction. This hints at an additional systematic, that we investigate in the following section.

\subsubsection{Examining the magnification of SN\,2022riv}\label{txt:sn2022riv}

\begin{figure}
\includegraphics[trim={0cm 0cm 0cm 0cm},clip,width=0.49\textwidth]{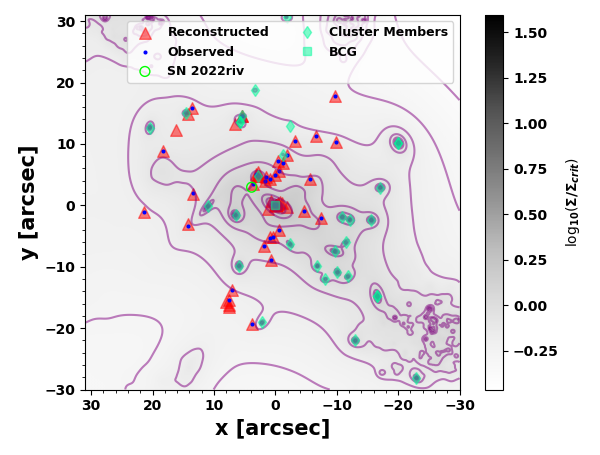}\\
\includegraphics[trim={0cm 0cm 0cm 0cm},clip,width=0.49\textwidth]{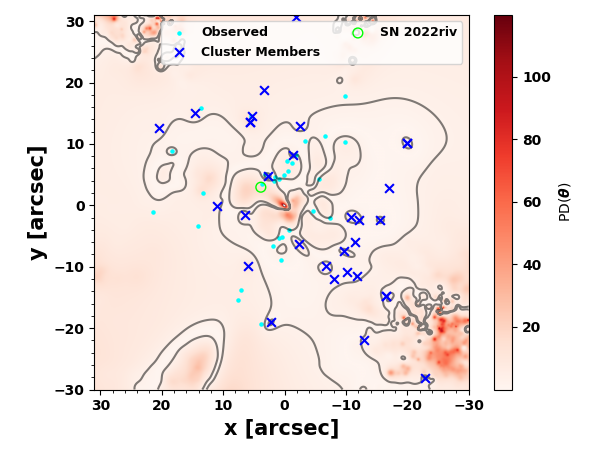}
\caption{\textit{Top:} Same as right panel of Figure \ref{fig:RXJ2129} but for the amended {\tt HoliGRALE} model that explicitly includes the BCG. \textit{Bottom:} Same as Figure \ref{fig:pdiffmaps} but for PD$\left(\boldsymbol{\theta}\right)=3$ (equation \ref{eq:PDmod}) between the original and updated RXJ2129 {\tt HoliGRALE} models.}
\label{fig:RXJ2129wbcg}
\end{figure}

The culprit for the magnification disagreement in our model is the small predicted shear ($\gamma$) at the SN location. This is somewhat surprising, as our model is consistent with the entire lens modeling sample of \cite{dhanasingham26} in many other predictions, such as the mass within 50 kpc ($M(r<50) \approx 1.7\times 10^{13} M_{\rm \odot}$), reconstructed image positions, central density profile slope, and time delays\footnote{SN\,2022riv is the last arriving image, and time delays have not been measured for its other two images. Therefore, consistency among the models in its prediction is not necessarily indicative of the accuracy of the models.}.

The reason for our model's small $\gamma$ is due to a systematic in our treatment of the BCG. In the blind analysis, we did not include the BCG as a constraint in the {\tt HoliGRALE} basis set. This was due to a modeling assumption of allowing the Plummers the freedom to reconstruct the BCG. From Figure \ref{fig:RXJ2129} and the examination of the model predictions, it does not at first appear to be a major systematic, as the freeform Plummer components of {\tt HoliGRALE} seem to compensate. However, the associated mass with the BCG is displaced by $\sim1.5''$ from the expected position away from SN\,2022riv. This displacement is what contributes to the large $\mu$ at SN\,2022riv.

We therefore amend the original {\tt HoliGRALE} model from \cite{dhanasingham26}. We redo the analysis with the same input parameters plus the explicit addition of a SIE component for the BCG. We initialize the BCG SIE with the results from \cite{jauzac21}. This model's density profile is shown in the top panel of Figure \ref{fig:RXJ2129wbcg}, and it achieves $\Delta_{\rm RMS}=0.05''$.

The amended model reconstructs the magnification of SN\,2022riv to be $-6.49\pm0.13$. This is now in much better agreement at $1.1\sigma$ with the observed magnification, and is consistent with the reconstructions in the original lens model sample.

We examine the inclusion of the BCG further in the bottom panel of Figure \ref{fig:RXJ2129wbcg}. Since the true mass distribution is unknown, we compare the updated and original model using the percent difference (PD) \citep{perera25b}:
\begin{equation}
    {\rm PD}(\boldsymbol{\theta}) = 2\frac{|\kappa_1(\boldsymbol{\theta}) - \kappa_2(\boldsymbol{\theta})|}{\kappa_1(\boldsymbol{\theta}) + \kappa_2(\boldsymbol{\theta})} \times 100 \label{eq:PDmod}
\end{equation}
where $\kappa_1$ and $\kappa_2$ are the two models being compared. We compute PD$\left(\boldsymbol{\theta}\right)$ between the original and updated RXJ2129 {\tt HoliGRALE} models, with the total MPD being 5.05\%. As can be seen, the models are very similar throughout, with the notable exception close to the BCG. Interestingly, the percent difference at the location of SN\,2022riv is 12.5\%, which is a much greater change than that of the entire cluster. This confirms that the inclusion of the BCG was integral to the reconstruction of the magnification.

There are two main takeaways from this analysis. The first is obvious: one should include all cluster members within the image window in order to ensure accurate reconstruction of lensing quantities. This represents a dramatic demonstration of the systematic bias when this is not taken into account. The second is that {\tt HoliGRALE} is able to adequately reconstruct magnifications, even for systems with low numbers of images such as RXJ2129. 

\subsubsection{Uncertainties and Comparison with {\tt GRALE}}\label{txt:RXJ2129uncertainties}

\begin{table}
\centering
\caption{SN 2022riv Magnification Reconstructions}
\label{tab:muSN}
\begin{tabular}{@{}lccc@{}}
\hline
 & Bootstrapped $\mu$ & Ensemble Averaged $\mu$ & $\Delta_{\rm RMS}$ \\
\hline
Original {\tt HoliGRALE} & $-15.39\pm0.85$ & $-15.93\pm7.17$ & $0.09''$ \\
Updated {\tt HoliGRALE} & $-6.49\pm0.13$ & $-6.48\pm0.75$ & $0.05''$ \\
FF {\tt GRALE} & $75.68\pm413.45$ & $2.78\pm37.10$ & $0.01''$ \\
\hline
\end{tabular}\\
\medskip{Summary of reconstructed $\mu$ of SN 2022riv from the three models made in this work. Section \ref{txt:uncertainty} describes the definitions of the Bootstrapped and Ensemble Averaged uncertainties. Original {\tt HoliGRALE} first appeared in \cite{dhanasingham26}, and the Updated {\tt HoliGRALE} model uses the same input but including the BCG explicitly. The measured $\mu = -5.35\pm1.01$. As can be seen, the Updated {\tt HoliGRALE} model is the only one that reconstructs the observed $\mu$ with meaningful precision.}
\end{table}

The uncertainties quoted previously regarding the $\mu$ of SN\,2022riv are computed via bootstrapping the {\tt HoliGRALE} ensemble as described in Section \ref{txt:uncertainty}. As mentioned in that section, this best quantifies the uncertainty of the complete (averaged) model, which is what is done in practice. However, as we described, the better way to quantify exactly how well {\tt HoliGRALE} is constraining a quantity is to examine the variance in the measured lensing observable ($\mu$ in this case) from the entire ensemble. We will do this for the two {\tt HoliGRALE} models mentioned in this section \citep[the original from][and the amended model that includes the BCG]{dhanasingham26} and for a third model that uses freeform {\tt GRALE}. These results are summarized in Table \ref{tab:muSN}.

The freeform {\tt GRALE} model is constructed in the typical way (see Section \ref{txt:grale}), and uses the same input image positions as the two {\tt HoliGRALE} models. This model achieves $\Delta_{\rm RMS}=0.01''$, considerably better than either {\tt HoliGRALE} model. However, the reconstructed magnification of SN\,2022riv (via bootstrapping) is $75.68\pm413.45$. It is therefore immediately apparent that the freeform {\tt GRALE} model is unable to constrain the SN\,2022riv $\mu$ with any meaningful precision. 

To examine this further, we compute the ensemble averaged $\mu$ from the 40 runs used to construct each model. For the original and updated {\tt HoliGRALE} models, $\mu$ is found to be $-15.93\pm7.17$ and $-6.48\pm0.75$, respectively. In both cases, the measured $\mu$ is very similar to the bootstrapped case. However, the fractional uncertainty is significantly increased in the Original model, rising from 6\% to 45\%. For the Updated model the rise is more muted, from 2\% to 12\%. For these reasons, we conclude that the Updated {\tt HoliRGRALE} model is a more well constrained model that is more accurately and precisely constraining $\mu$.

The freeform {\tt GRALE} model recovers $\mu=2.78\pm37.10$. In addition to being an imprecise reconstruction, the fractional uncertainty massively increases. Furthermore, the large disagreement with the bootstrapped case emphasizes the poor constraining power of the freeform model. From this, we conclude that freeform {\tt GRALE} is an under-constrained and imprecise model. This is interesting because the $\Delta_{\rm RMS}$ is the lowest of the three models considered. We interpret this as indicating that the freeform {\tt GRALE} model is overfit. This then serves as a case study in the pitfalls of relying on $\Delta_{\rm RMS}$ as an indicator of lens model accuracy. In particular for this case, the only reason {\tt GRALE} performs poorly is due to the low number of image positions. Likewise, this is consistent with the results from synthetic clusters shown in Section \ref{txt:synthetic}, where low $\Delta_{\rm RMS}$ associated with small $N_{\rm im}$ does not correlate with more accurate lens models.

\subsection{MACS J0138.0$-$2155}\label{txt:MACSJ0138}

\begin{figure}
\includegraphics[trim={0cm 0cm 0cm 0cm},clip,width=0.49\textwidth]{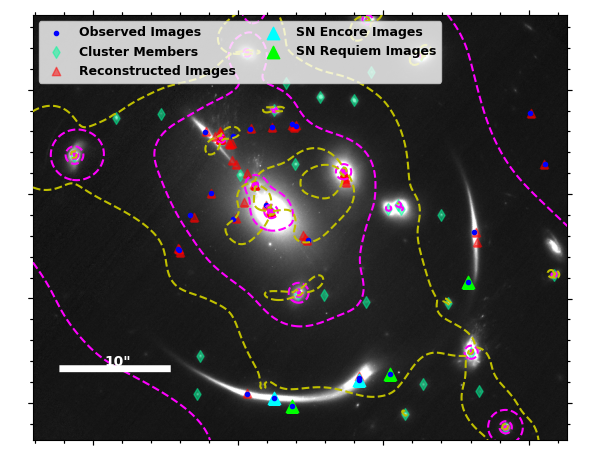}
\caption{JWST F200W view of MACSJ0138. Overplotted magenta and yellow contours indicate the HG\_wTD\_MP model density distribution and critical curve for $z=1.949$, respectively. Cluster members, reconstructed images, and observed images are shown as green diamonds, red triangles, and blue dots, respectively. Cyan and bright green triangles show the multiple images of SN Encore and SN Requiem, respectively.}
\label{fig:MACSJ0138}
\end{figure}

\subsubsection{Reconstruction of Observed Time Delays}\label{txt:tdrecon}

\begin{table*}
\centering
\caption{MACSJ0138 Lens Model Results}
\label{tab:MACSJ0138}
\setlength{\tabcolsep}{6pt}
\begin{tabular}{@{}cccccc@{}}
\hline
& FF & HG\_noTD\_SP & HG\_noTD\_MP & HG\_wTD\_SP & HG\_wTD\_MP \\
\hline
$\Delta_{\rm RMS}$ & $0.08''$ & $0.13''$ & $0.11''$ & $0.20''$ & $0.21''$ \\
Bootstrapped $\Delta\tau_{\rm 1b,1a}$ & $-0.65^{+6.64}_{-6.60}$ (5.20$\sigma$) & $-15.0^{+9.34}_{-14.59}$ (1.98$\sigma$) &
$-14.59^{+4.04}_{-5.46}$ (4.23$\sigma$) & 
$-21.84^{+5.32}_{-7.21}$ (2.49$\sigma$) &
$-26.01^{+7.24}_{-9.22}$ (1.54$\sigma$) \\
Ensemble Averaged $\Delta\tau_{\rm 1b,1a}$ & $-7.96^{+54.42}_{-64.93}$ (0.53$\sigma$) & $-17.22^{+13.46}_{-17.61}$ (1.42$\sigma$) & 
$-20.72^{+19.5}_{-26.62}$ (0.82$\sigma$) &
$-31.79^{+12.02}_{-8.86}$ (0.73$\sigma$) &
$-33.92^{+15.61}_{-14.62}$ (0.38$\sigma$) \\
Bootstrapped $\Delta\tau_{\rm 2a,2b}$ & $37.37^{+4.72}_{-5.37}$ (2.56$\sigma$) & $44.12^{+6.69}_{-4.97}$ (2.32$\sigma$) & $51.18^{+6.12}_{-5.12}$ (2.09$\sigma$) & $78.13^{+5.45}_{-7.30}$ (1.19$\sigma$) & 
$69.47^{+5.24}_{-5.34}$ (1.49$\sigma$) \\
Ensemble Averaged $\Delta\tau_{\rm 2a,2b}$ & $38.92^{+26.79}_{-31.86}$ (1.81$\sigma$) & $44.67^{+16.18}_{-18.77}$ (2.02$\sigma$) & 
$48.0^{+17.92}_{-19.03}$ (1.90$\sigma$) & 
$70.86^{+24.32}_{-37.79}$ (1.01$\sigma$) &
$71.08^{+22.76}_{-34.07}$ (1.05$\sigma$) \\
Bootstrapped $\Delta\tau_{\rm 2c,2b}$ & $-157.83^{+24.14}_{-21.93}$ (4.87$\sigma$) & $-57.41^{+16.61}_{-15.53}$ (1.70$\sigma$) & $-43.78^{+16.06}_{-19.22}$ (1.08$\sigma$) & 
$-34.53^{+8.03}_{-9.59}$ (0.89$\sigma$) &
$-31.06^{+6.20}_{-5.47}$ (0.76$\sigma$) \\
Ensemble Averaged $\Delta\tau_{\rm 2c,2b}$ & $-163.48^{+135.98}_{-136.01}$ (1.07$\sigma$) & $-64.29^{+62.64}_{-67.84}$ (0.70$\sigma$) &  
$-51.15^{+93.02}_{-72.90}$ (0.40$\sigma$) & 
$-37.25^{+18.48}_{-28.65}$ (0.69$\sigma$) &
$-41.18^{+16.73}_{-18.03}$ (0.98$\sigma$)\\
\hline
\end{tabular}\\
\medskip{Summary of lens modeling results for our five models of MACSJ0138. SN\,Encore's time delay is measured to be $\Delta\tau_{\rm 1b,1a} = -39.8^{+3.9}_{-3.3}$ days \citep{pierel26}, while two time delays are measured for SN\,Requiem: $\Delta\tau_{\rm 2a,2b} = 114^{+28}_{-31}$ days and $\Delta\tau_{\rm 2c,2b} =  -17^{+19}_{-16}$ days \citep{rodney21}. Results presented here show the model predicted time delays for these supernovae. Uncertainties represent the 68\% confidence interval. Parenthesis indicate $N_{\sigma}$ for the observation (equation \ref{eq:numsigma}).}
\end{table*}

MACS J0138.0$-$2155 (hereafter MACSJ0138) is the subject of much interest thanks to the discovery to two lensed supernovae within the same source galaxy ($z_{\rm s} = 1.949$): SN\,Requiem \citep{rodney21} and SN\,Encore \citep{pierel24}. As a result, numerous lens models have been developed \citep{acebron25,ertl25,bazzanini26}, which has led to constraints on dark matter \citep{odonnell26} and the Hubble Constant ($H_0$) \citep{suyu26,pierel26}. The latter effort is possible thanks to time delay measurements from both of these SNe. This makes MACSJ0138 an ideal testing ground for {\tt HoliGRALE} systematics on constraining time delays. We show the MACSJ0138 field and one of the lens models used in this study in Figure \ref{fig:MACSJ0138}.

MACSJ0138 ($z_{\rm d}=0.336$) contains 23 multiple images from 8 systems across 4 spectroscopically confirmed source redshift planes \citep{ertl25}. It also contains numerous cluster member galaxies spectroscopically identified with MUSE \citep{granata25}. For our {\tt HoliGRALE} models, we only include those cluster members from \cite{granata25} that lie within our modeling window of $100''\times100''$ centered on the BCG. Cluster members are modeled as SIEs with velocity dispersions initialized from stellar kinematics measurements with MUSE. Cluster members lacking measured velocity dispersions are initialized according to the derived Faber-Jackson relation from \cite{granata25}. In addition, we also include the three jellyfish galaxies, initialized with their observed stellar velocity dispersions \citep{gibson26}. Furthermore, we consider the effect of two line-of-sight galaxies, one foreground ($z=0.3087$) and one background ($z=0.371$). The foreground galaxy (J4 in \cite{gibson26}) lies very close to the SNe images, while the background galaxy (ID 82 in Table 1 of \cite{granata25}) is measured to have a large mass \citep{flowers25}. Since the {\tt GRALE} (and {\tt HoliGRALE} by extension) multi-plane lens modeling paradigm is still experimental, we include them in our models by placing them at the lens redshift and rescaling their SIE component by their angular diameter distance. Since the foreground and background redshifts are very close to the lens plane redshift, and since {\tt HoliGRALE} will rescale the weights of these components, we justify this modeling approximation as sufficient to first order.

To rigorously examine the accuracy of constraining time delays with {\tt HoliGRALE}, we build 5 models, all using the same 23 input images. The first model is built with freeform {\tt GRALE} (FF), and serves as our control. The second and third models (HG\_noTD\_SP and HG\_noTD\_MP, respectively) are the baseline {\tt HoliGRALE} models, which include all cluster members mentioned. Both models do not include time delay information in the fitness criterion (see Section \ref{txt:fitness}). ``SP'' and ``MP'' indicate single plane (exclusion of the foreground and background galaxies) and multi-plane (inclusion of the foreground and background galaxies). The fourth and fifth models (HG\_wTD\_SP and HG\_wTD\_MP, respectively) are the same as HG\_noTD\_SP and HG\_noTD\_MP except with time delays explicitly included in the fitness criterion as constraints. Figure \ref{fig:MACSJ0138} shows the HG\_wTD\_MP model as a sample result.

There are three observed time delays in MACSJ0138. SN\,Encore (System 1 in \citealt{ertl25}) has only two observed images (1a and 1b), with a time delay\footnote{The convention we use for reporting time delays is $\Delta\tau_{ij}=\tau_i - \tau_j$, which indicates the time delay of image $i$ with respect to image $j$.} $\Delta\tau_{\rm 1b,1a} = -39.8^{+3.9}_{-3.3}$ days \citep{pierel26}. The time delay is well constrained for SN Encore due to being spectroscopically classified as a Type Ia with multiple epochs of observations. SN\,Requiem (System 2 in \citealt{ertl25}) has three observed images (2a, 2b, and 2c), with two measured time delays \citep{rodney21}: $\Delta\tau_{\rm 2a,2b} = 114^{+28}_{-31}$ days and $\Delta\tau_{\rm 2c,2b} =  -17^{+19}_{-16}$ days. These time delays are not as well constrained since SN\,Requiem has only one epoch of photometry and is assumed to be an SN Ia. HG\_wTD\_SP and HG\_wTD\_MP uses these as constraints. We emphasize that our analysis is not making an attempt to measure $H_0$, although in principle it can be done using our FF, HG\_noTD\_SP, and HG\_noTD\_MP models. This analysis is instead attempting to understand the systematics in measuring time delays with {\tt HoliGRALE}.

We note that the time delays reported in \cite{rodney21} use lens model magnifications to constrain $\Delta\tau_{\rm 2a,2b}$ and $\Delta\tau_{\rm 2c,2b}$. Measured time delays not based on lens mass models are $\Delta\tau_{\rm 2a,2b} = 116.2^{+29.0}_{-29.9}$ days and $\Delta\tau_{\rm 2c,2b} =  1.3^{+32.5}_{-32.9}$ days. These are the time delays adopted in \cite{suyu26} and \cite{bazzanini26}, although their models do not directly use them as constraints in the lens modeling. For this work, we do not adopt the non-lens model time delay measurements due to them favoring a possibly nonphysical image configuration. The host galaxy source (System 3 in \citealt{ertl25}) of both supernovae is a quadruply imaged system (quad) with an observed central image. According to gravitational lensing theory, one can observe the positions of images in quads and estimate the approximate arrival time order of the 5 images from their morphology \citep{saha03}. Applying this methodology to MACSJ0138 results in image 3c being the first arriving image. Therefore, $\Delta\tau_{\rm 2c,2b}<0$. While the model-independent time delay measurement is consistent with this, its posterior peaks in the flipped time delay configuration and is broadly consistent with a positive time delay. Whereas, the model dependent $\Delta\tau_{\rm 2c,2b} =  -17^{+19}_{-16}$ days more strictly abides by the theoretical constraint. We note that it is technically possible to flip the arrival time order for a cusp lens (which MACSJ0138 is approximately) if there is mass perturber nearby the first arriving image \citep[][for a galaxy-scale example]{keeton09}, although this would be an exceptional case. Furthermore, the large uncertainty for the SN\,Requiem time delays renders this a subdominant constraint in the lens model. 

The results from our models are shown in Table \ref{tab:MACSJ0138}. As in Section \ref{txt:RXJ2129}, we report both the bootstrapped and ensemble averaged uncertainties. For each time delay reconstruction, we compute the number of $\sigma$ ($N_{\sigma}$) each reconstruction is from the observed time delay, with $\sigma$ added in quadrature:
\begin{equation}
    N_{\sigma} = \frac{|\Delta\tau_{\rm mod} - \Delta\tau_{\rm obs}|}{\sqrt{\sigma_{\rm mod}^2 + \sigma_{\rm obs}^2}}\label{eq:numsigma}
\end{equation}
$\sigma_{\rm mod}$ and $\sigma_{\rm obs}$ are taken to be the model and observed mean 68\% confidence interval as reported. 

As can be seen, the ensemble averaged measurements from all models are able to constrain each observed time delay to $\lesssim2\sigma$. The FF model exhibits a large fractional uncertainty, allowing us to conclude that this model is under-constraining time delays. Like RXJ2129, this is not so surprising due to the small number of observed image positions. In general, the inclusion of the three time delays as model constraints results in a more accurate and precise reconstruction of time delays. This indicates that time delay information, when available, is an important driving factor in improving the accuracy of {\tt HoliGRALE}. This conclusion is drawn from the ensemble averaged uncertainties, which fully accounts for the variability of {\tt HoliGRALE}. We note that this trend remains true for all the bootstrapped mean uncertainties as well, indicating that our full complete model, as would be used in a sample, are also improved upon inclusion of time delays. 

For SN\,Requiem's time delays, the observed time delays are not well constrained. Despite this, the models used in \cite{suyu26} exhibit substantial disagreement with $\Delta\tau_{\rm 2a,2b}$. All models used in their sample, except LenstoolI, are in $>1.5\sigma$ disagreement with the observation. These models, importantly, do not directly use time delays as constraints. We find that for our models that do not include $\Delta\tau_{\rm 2a,2b}$ as a constraint (FF, HG\_noTD\_SP, and HG\_noTD\_MP), this systematic persists, with these models exhibiting $\sim2\sigma$ disagreement. Including time delays, however, reduces this disagreement to $\sim1\sigma$. We note that the underlying tension might be due to systematics associated with the $\Delta\tau_{\rm 2a,2b}$ measurement, and not in the lens modeling \citep{suyu26}.

For SN\,Encore, $\Delta\tau_{\rm 1b,1a}$ is much more precisely measured \citep{pierel26}. Our best model at reconstructing $\Delta\tau_{\rm 1b,1a}$, HG\_wTD\_MP, reconstructs it to $0.38\sigma$. In this case, there exists a systematic when computing the complete (averaged) model from the ensemble, as the bootstrapped $\Delta\tau_{1\rm b,1a}$ is substantially shorter (and in $1.54\sigma$ disagreement) than the ensemble averaged prediction. Following our discussion in Section \ref{txt:RXJ2129uncertainties}, we mentioned that a measure of how well constrained a model is can be determined qualitatively by comparing the bootstrapped and ensemble averaged measurements. In the context for SN\,Encore, this hints that the HG\_wTD\_MP and HG\_wTD\_SP models are not as well constrained. Despite this, they remain a substantial improvement in time delay prediction over the three models that do not include $\Delta\tau_{\rm 1b,1a}$ as a constraint.

\subsubsection{Time Delay Predictions}\label{txt:tdpred}

\begin{figure}
\includegraphics[trim={0cm 0cm 0cm 0cm},clip,width=0.49\textwidth]{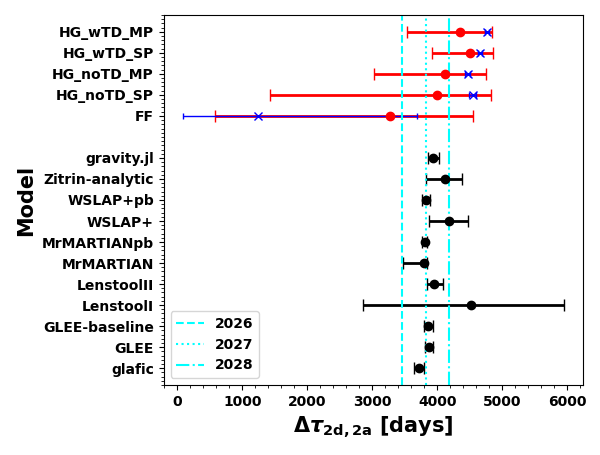}
\caption{Predicted time delays for the 4th arriving image of SN\,Requiem ($\Delta\tau_{\rm 2d,2a}$). Models in red are the 5 models presented in this work (see Section \ref{txt:tdrecon}), with ensemble averaged errors. The bootstrapped results are shown in blue. Models in black are the lens models used in the \cite{suyu26} sample and the {\tt gravity.jl} model from \cite{bazzanini26}. Cyan lines indicate January 1 of the respective year as a reference point.}
\label{fig:tdelays}
\end{figure}

\begin{figure}
\includegraphics[trim={0cm 0cm 0cm 0cm},clip,width=0.49\textwidth]{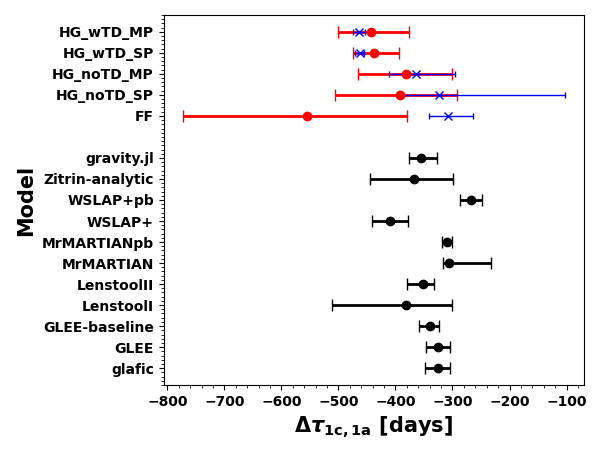}
\includegraphics[trim={0cm 0cm 0cm 0cm},clip,width=0.49\textwidth]{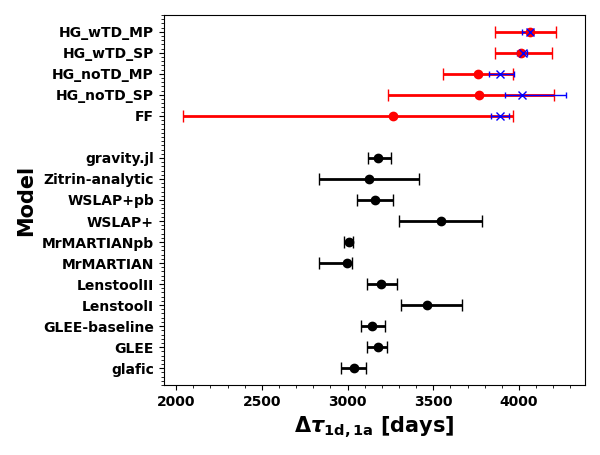}
\includegraphics[trim={0cm 0cm 0cm 0cm},clip,width=0.49\textwidth]{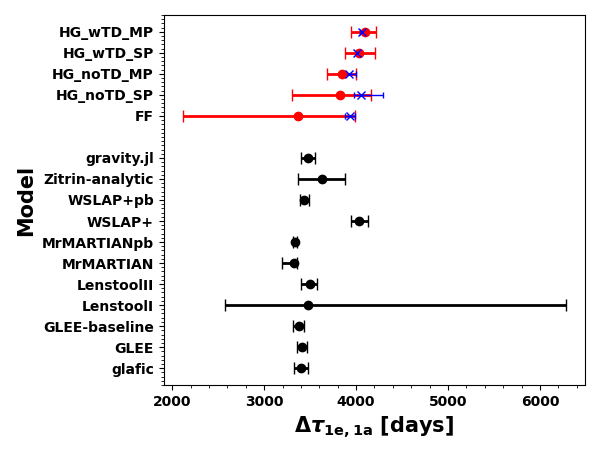}
\caption{Same as Figure \ref{fig:tdelays} but for the 1st ($\Delta\tau_{\rm 1c,1a}$), 4th ($\Delta\tau_{\rm 1d,1a}$), and 5th ($\Delta\tau_{\rm 1e,1a}$) arriving image of SN\,Encore.}
\label{fig:tdelaysEncore}
\end{figure}

At the time of this writing, the fourth image of SN\,Requiem is expected to arrive in 2027, and is expected to provide a $\sim$2--3\% measurement of $H_0$. For this reason, we present our model predictions for the time delay of the fourth arriving image of SN\,Requiem ($\Delta\tau_{\rm 2d,2a}$). Our results, along with a comparison with previous results, is shown in Figure \ref{fig:tdelays}. Full SN\,Requiem time delay constraints are presented in Table \ref{tab:SNRequiemTDpred} in Appendix \ref{txt:encoreTDpreds}. Our most informed model, HG\_wTD\_MP, predictes $\Delta\tau_{\rm 2d,2a} = 4346.61^{+490.27}_{-818.36}$ days.

As can be seen, the uncertainty on the predicted $\Delta\tau_{\rm 2d,2a}$ significantly reduces when time delays are included in the model. All of our models are broadly consistent with previous results. FF exhibits the largest uncertainty. Overall, all of our models show a wider ensemble averaged uncertainty limit than the models of \cite{suyu26} and \cite{bazzanini26}. Bootstrapped uncertainties show a more comparable fractional uncertainty, and are consistent to $1\sigma$ with the ensemble average indicating well constrained results. Our FF model is the only one that fails to precisely constrain $\Delta\tau_{\rm 2d,2a}$ with any useful precision.

An interesting result is that all our {\tt HoliGRALE} models imply a longer time delay than the predictions of \cite{suyu26} and \cite{bazzanini26}. Focusing on HG\_noTD\_SP and HG\_noTD\_MP, as these can be used to infer $H_0$, this would imply a lower $H_0$ result according to the analysis in Figure 10 of \cite{suyu26}. 

We present in Figure \ref{fig:tdelaysEncore} (and Table \ref{tab:SNEncoreTDpred} in Appendix \ref{txt:encoreTDpreds}) the time delay predictions for SN\,Encore. Likewise, our models are in broad agreement with those from \cite{suyu26} and \cite{bazzanini26}. For the 4th arriving image time delay ($\Delta\tau_{\rm 1d,1a}$), we similarly reconstruct a longer time delay prediction. Since this image forms very near to the 4th arriving SN\,Requiem image, we hypothesize that the reason for this systematic is due to the reconstructed mass of the J1 Jellyfish galaxy \citep{gibson26}. This is also found to be a large source of scatter in the lens reconstructions of \cite{suyu26}, and we also note that future image detections in this region (including the 4th arriving SN\,Requiem image) will place tighter constraints on its mass, which can be used to improve lens models.

The main takeaway from this discussion is that {\tt HoliGRALE} predicts future time delays with better precision when time delays are included. The uncertainties of full (mean) {\tt HoliGRALE} models are consistent with other methods, and the ensemble averaged results show that {\tt HoliGRALE} is effectively reconstructing the time delays. Our results for the 4th arriving SN\,Requiem image can be used in the near future as a prediction for when it is ultimately observed.

\subsection{Abell S1063}

\begin{figure}
\includegraphics[trim={0cm 0cm 0cm 0cm},clip,width=0.49\textwidth]{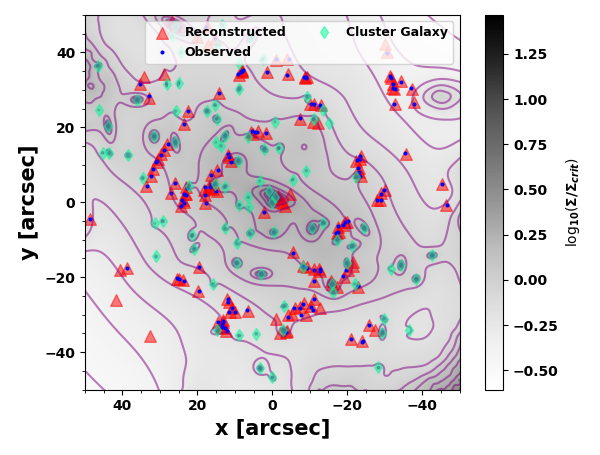}
\caption{Full density distribution of the {\tt HoliGRALE} model of AS1063, with purple density contours separated by $\Delta\log_{10}\kappa=0.1$. Cluster members, reconstructed images, and observed images are shown as green diamonds, red triangles, and blue dots, respectively.}
\label{fig:AS1063}
\end{figure}

Abell\,S1063 (hereafter AS1063) is the last system we will examine in this work. AS1063 has observations from the Hubble Frontier Fields \citep{lotz17}, BUFFALO survey \citep{steinhardt20}, and JWST's GLIMPSE program \citep{atek25}. It has been extensively modeled over the past decade \citep[e.g.,][]{diego16,caminha16,kawamata18,bergamini19,chan20,granata22,limousin22,cha23,beauchesne24,diego26a,beauchesne26} and has been used to constrain numerous things. In this section, we focus in particular on the consistently recovered cored density profile of AS1063 reconstructed by many independent lens models \citep{bergamini19,limousin22,granata22,beauchesne26,diego26b}. This result is intriguing as it is potentially in tension with CDM expectations, which prefer cuspy density profiles, although formulations of CDM theory from statistical mechanics can predict cores \citep{williams10,francis25}. A cored density profile is also consistent with self-interacting dark matter (SIDM) \citep{spergel00}, warranting further analysis of AS1063 to understand the nature of dark matter. 

We construct a {\tt HoliGRALE} lens model of AS1063 utilizing the HST/JWST image position catalog of \cite{diego26a} which consists of 126 spectroscopically confirmed images across 44 multiply imaged systems. Cluster member galaxies are initialized as SIEs with velocity dispersions measured in a two step process. First, effective radii measurements are gathered from \cite{tortorelli23}. Then, velocity dispersions are calculated from the Fundamental Plane \citep{djorgovski87}. The resulting model achieves $\Delta_{\rm RMS}=0.53''$, and is briefly shown in \cite{diego26a} for comparison purposes. We present the full lens model map in Figure \ref{fig:AS1063}.

To study the cored profile of AS1063, we compare {\tt HoliGRALE} with the hybrid {\tt WSLAP+} model from \cite{diego26a}, which uses the same image inputs as our model. We also include the Hubble Frontier Fields (HFF) freeform {\tt GRALE} model\footnote{This model can be accessed as the ``Williams'' model on the HFF site: \url{https://archive.stsci.edu/pub/hlsp/frontier/abells1063/models/}} ({\tt GRALE16}) for posterity. {\tt GRALE16} only uses HST data (50 images from 19 sources), so we do not directly compare it with {\tt HoliGRALE}. However, as we will see, its inclusion serves as a benchmark for the pre-JWST understanding of the AS1063 density profile. Importantly, a large core using the same image dataset as {\tt GRALE16} is reconstructed in parametric models \citep{bergamini19,granata22,limousin22}, so its inclusion helps to faciliate an apples-to-apples comparison with contemporary lens models.

\begin{figure}
\includegraphics[trim={0cm 0cm 0cm 0cm},clip,width=0.49\textwidth]{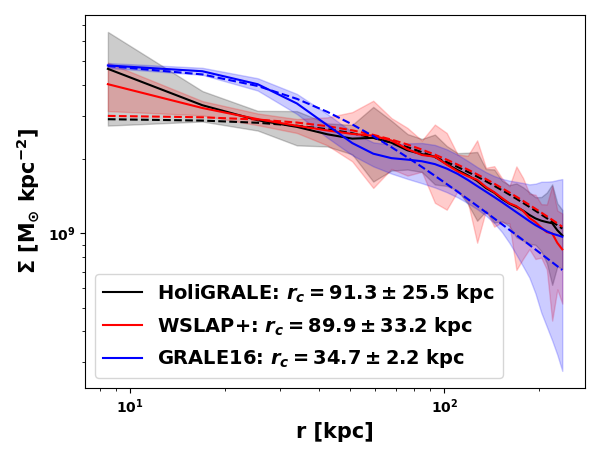}
\caption{Spherically averaged surface density ($\Sigma$) profiles for {\tt HoliGRALE} (black), {\tt WSLAP+} (red), and HFF {\tt GRALE} ({\tt GRALE16}; blue). Dashed lines are the best fit spherical dPIE (equation \ref{eq:dpie}) for each model. The core radius ($r_{\rm c}$) measured from this fit is presented.}
\label{fig:AS1063corerad}
\end{figure}

We present the spherically averaged density profiles of these three models in Figure \ref{fig:AS1063corerad}. To quantify the core radius, we follow \cite{granata22} and fit a spherical dual pseudo-isothermal elliptical (dPIE) profile \citep{eliasdottir07}:
\begin{equation}
    \Sigma(R) = \frac{\sigma_0^2}{2G}\frac{r_{\rm t}}{r_{\rm t} -r_{\rm c}}\left(\frac{1}{\sqrt{R^2 + r_{\rm c}^2}} - \frac{1}{\sqrt{R^2 + r_{\rm t}^2}} \right)\label{eq:dpie}
\end{equation}
where $\sigma_0$ is the central velocity dispersion, $r_{\rm t}$ is the truncation radius, and $r_{\rm c}$ is the core radius. \cite{bergamini19} recovers $r_{\rm c} = 88.6\pm0.53$\,kpc, \cite{granata22} recovers $r_{\rm c}=86\pm2$\,kpc, and \cite{limousin22} recovers $r_{\rm c}=89.5\pm5.5$\,kpc using spherical dPIE fits to their models. All of these models use parametric {\tt LensTool} and are in excellent agreement with one another. As seen in Figure \ref{fig:AS1063corerad}, {\tt GRALE16}, which uses the same input image dataset, does not recover a large core, with $r_{\rm c}=34.7\pm2.2$\,kpc. So pre-JWST contemporary models showed disagreement in the existence of a core. We note that the core radius depends on the metric used to measure it, and some studies recover a core in AS1063 using different central profiles \citep[for example, a Burkert profile is used in][]{diego26b}.

JWST imaging has doubled the number of images for AS1063. With this new data, \cite{beauchesne26} reports $r_{\rm c}=99.9\pm3.0$\,kpc using {\tt LensTool}, in reasonable agreement and consistency with previous {\tt LensTool} results. As seen in Figure \ref{fig:AS1063corerad}, {\tt HoliGRALE} reconstructs $r_{\rm c}=91.3\pm25.5$\,kpc, also in agreement with previous results. {\tt HoliGRALE} is in excellent agreement with {\tt WSLAP+}, for which we measure $r_{\rm c}=89.9\pm33.2$\,kpc. The larger uncertainty in these two cases is due to a different method of calculation; uncertainties on the fit are from Monte Carlo errors, rather than an MCMC posterior as in the case of {\tt LensTool}. Our results substantiate the claim from many previous works that AS1063 exhibits a core of $\sim90$\,kpc.

We note that the existence of a core in AS1063 is an ongoing research question. While lens modeling results seem to indicate a large core, this could be the result of a hidden systematic. This scenario was tested in \cite{limousin22}, where they found that non-cored lens models result in a significant increase in $\Delta_{\rm RMS}$ and unphysically large velocity dispersion predictions. They conclude that the most physically realistic priors are most consistent with a core. For non-parametric lensing, such as hybrid models {\tt HoliGRALE} and {\tt WSLAP+}, this is not an issue as the shape of the dark matter density profile is not specified a priori. Therefore, the consistent recovery of a large core across all classes of lens models provides stronger support to the core's existence. \cite{diego26b} goes a step further and notes that the large core is also seen in the globular cluster distribution, a model independent observation. They note the strengths and weaknesses of the CDM and SIDM explanations for the core, and conclude that higher resolution hydrodynamical simulations are necessary to differentiate the two dark matter models. 

From the perspective of {\tt HoliGRALE}, its consistency with many of its contemporary lens models of AS1063 in reconstructing a large core is encouraging. {\tt HoliGRALE} appears to join a growing agreement of a large $\sim90$\,kpc core in AS1063, which has implications on the nature of dark matter. Interestingly, freeform {\tt GRALE}, when compared with its contemporary lens models, was in disagreement on the core's existence, although it is in $<2\sigma$ agreement with {\tt HoliGRALE}. Future observations and simulations will shed light on the existence of this core, potentially illuminating the nature of dark matter.

\section{Discussion and Conclusions}\label{txt:conclusions}

We present a thorough examination of {\tt HoliGRALE}, a hybrid extension of the freeform lens inversion algorithm, {\tt GRALE}. The goal of this paper is to understand the systematics of {\tt HoliGRALE}, whose primary difference is the inclusion of parametric components for cluster members in its underlying freeform genetic algorithm. We test these against three synthetic clusters from IllustrisTNG of varying complexity, and analyze the performance of {\tt HoliGRALE} in terms of its reconstruction of the density distribution, image positions, magnifications, and time delays. We also use {\tt HoliGRALE} to model three real cluster lens systems, and evaluate its performance on real data and in comparison with other lens modeling codes. Our main results are as follows:
\begin{itemize}
    \item For simple and small clusters, {\tt HoliGRALE} outperforms {\tt GRALE} in reconstructing the density distributions. We expect the uncertainty budget in the entire lens to be $6.1\pm1.0\%$. For more complex clusters with many cluster member galaxies, {\tt HoliGRALE} is on par with {\tt GRALE}, with an expected error budget of $\sim$10--20\%. In the inner regions of clusters, where there are more images, {\tt HoliGRALE} is an improvement over {\tt GRALE}, with an error budget of $\sim$3--10\%. We confirm that the driving factor of the improvement with {\tt HoliGRALE} is the inclusion of cluster members explicitly in the genetic algorithm.
    \item In reconstructing image positions, we find that the failure modes for reconstructing certain image positions are dominated by small scale structure. We confirm the degree of this effect by computing the cross-correlation function between the input image position number density and the percent difference map with the true density distribution. We find that {\tt HoliGRALE} is able to more accurately reconstruct the density distribution in regions with high image number density. We therefore expect reliable constraints on the density distribution to be primarily in regions with many observed images. Furthermore, magnifications are very well reconstructed in {\tt HoliGRALE}. Time delays are well reconstructed in simple clusters, and moderately well reconstructed in clusters with more complex mass features.
    \item After validation of {\tt HoliGRALE} on synthetic clusters, we apply it to three presently applicable lenses of interest. In RX\,J2129.7$+$0005, we analyze the observed magnification of SN\,2022riv, finding that {\tt HoliGRALE} can most accurately reconstruct it only when incorporating all cluster members in its analysis, despite the lens's low image number density. In MACS J0138.0$-$2155, {\tt HoliGRALE} exhibits substantial improvement in its recovery of observed time delays once they are taken into account. We use our models to predict the upcoming time delay of the fourth arriving image of SN\,Requiem to be $4346.61^{+490.27}_{-818.36}$\,days. In Abell\,S1063, {\tt HoliGRALE} reconstructs a large core with a core radius of $r_{\rm c}=89.9\pm33.2$\,kpc, consistent with the results of many contemporary models and can be used to constrain alternative theories of dark matter.
\end{itemize}

In the current landscape of cluster lens modeling, rigorous stress testing and understanding of underlying systematics is essential in order to trust predictions. This work serves that purpose for {\tt HoliGRALE}, and we are able to quantify the degree of accuracy with which we expect to reconstruct lenses of varying complexity. Future studies utilizing {\tt HoliGRALE} models will have a better understanding of the systematics associated with their measurements.

Further improvements to {\tt HoliGRALE} include testing more profiles to incorporate into the genetic algorithm, understanding the potential benefits of including weak lensing information similar to \cite{liesenborgs20}, and testing the multi-plane lensing approach. In the context of the field of cluster lens modeling, another interesting direction would be to move beyond traditional position-based modeling and into the extended surface brightness modeling regime. Much progress has been made on this topic recently \citep[e.g.,][]{acebron24, eid26, schuldt26}, and it is worthwhile to explore.

Lastly, lensing degeneracies \citep[e.g.,][]{williams26b} continue to plague lens models \citep{perera25b}. A generalized understanding of lensing degeneracies would be ideal in order to better understand the systematics in measurements from gravitational lenses. It is therefore important to scrutinize the predictions of lens models thoroughly. This has been done in the past with {\tt GRALE} \citep{liesenborgs08b, liesenborgs12, sebesta19, ghosh23, liesenborgs24, limousin25} and was demonstrated in this work as a first step for three systems. This lays the groundwork for future studies with {\tt HoliGRALE} to continue this precedent. 

Hybrid lens modeling has long been advocated as an important future step in improving mass reconstructions of clusters \citep{meneghetti17}. The results of this work quantify the robustness of the hybrid lens models of {\tt HoliGRALE} against simulated clusters and present three new models of real clusters with accurate reconstructions, establishing its effectiveness for future studies.

\section*{Acknowledgements}
The authors acknowledge the computational resources provided by the Minnesota Supercomputing Institute (MSI), which were essential for this study. 

\section*{Data Availability}
All data, lens models, and tutorial examples from this paper will be made publicly available upon acceptance of this paper.

\bibliographystyle{mnras}
\bibliography{references}


\appendix

\section{Appendix A: Mass Model Maps for Synthetic Clusters}\label{txt:pdiffmaps}
\begin{figure*}
\includegraphics[trim={0cm 0cm 0cm 0cm},clip,width=0.49\textwidth]{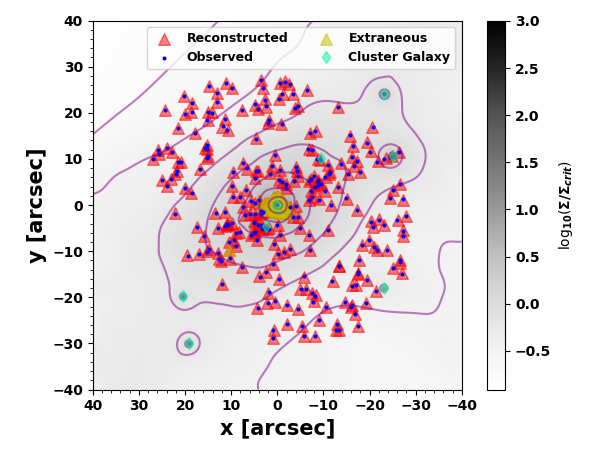}
\includegraphics[trim={0cm 0cm 0cm 0cm},clip,width=0.49\textwidth]{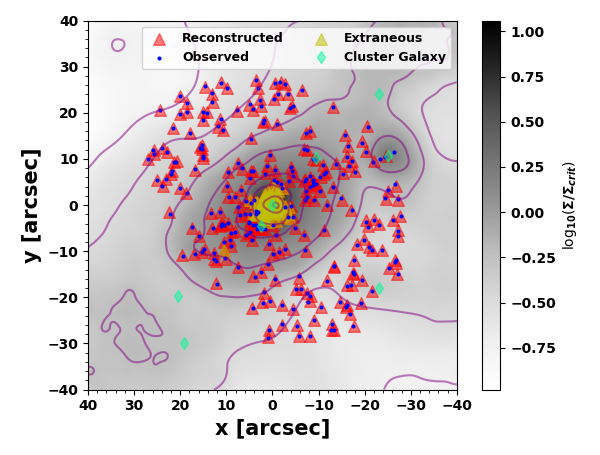}\\
\includegraphics[trim={0cm 0cm 0cm 0cm},clip,width=0.49\textwidth]{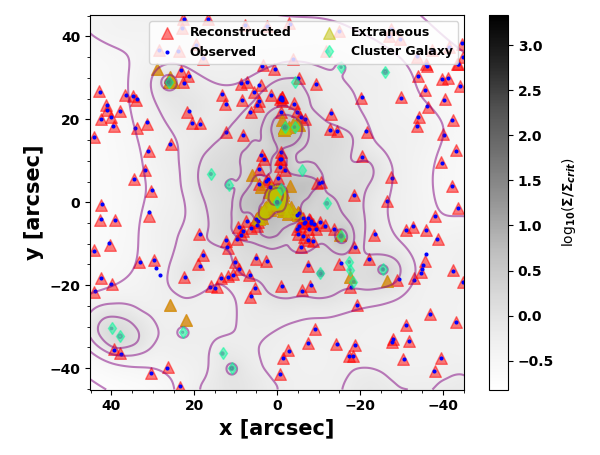}
\includegraphics[trim={0cm 0cm 0cm 0cm},clip,width=0.49\textwidth]{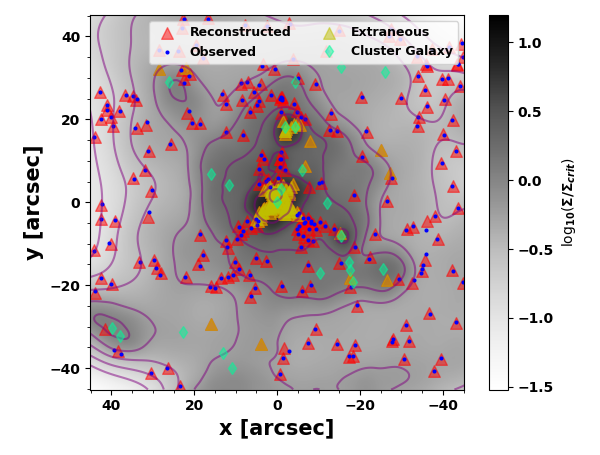}\\
\includegraphics[trim={0cm 0cm 0cm 0cm},clip,width=0.49\textwidth]{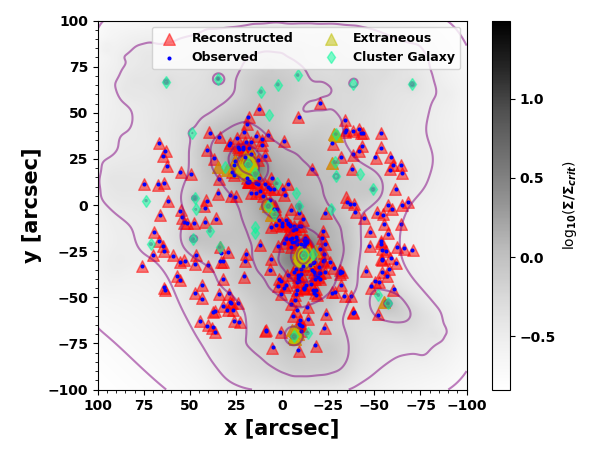}
\includegraphics[trim={0cm 0cm 0cm 0cm},clip,width=0.49\textwidth]{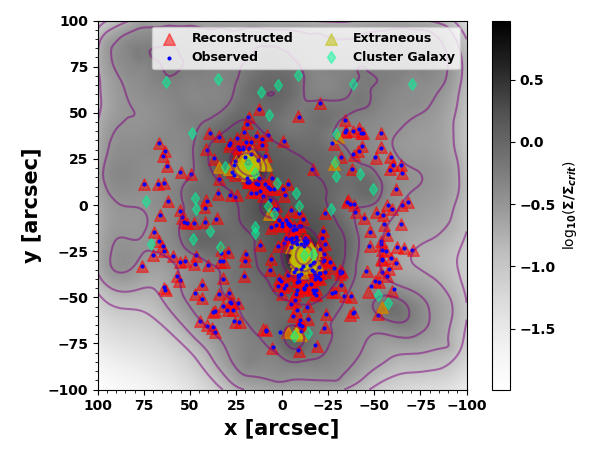}\\
\caption{Reconstructed density distributions for the $N_{\rm S}=100$ {\tt HoliGRALE} (\textit{Left Column}) and {\tt GRALE} (\textit{Right Column}) models of Harriet (\textit{Top}), Maka Ska (\textit{Middle}), and Nokomis (\textit{Bottom}). Observed images used as input are shown as blue dots. Reconstructed images from the model are overplotted as red triangles. Green diamonds indicate cluster member galaxies. Extraneous images (including central demagnified unobservable images) are indicated as yellow triangles. $\Sigma_{\rm crit}$ is computed for $D_{\rm ds}/D_{\rm s} = 1$.}
\label{fig:massreconmaps}
\end{figure*}

In Section \ref{txt:massrecon}, the accuracy of the mass reconstructions is evaluated on the synthetic clusters. Results are quoted in terms of the MPE (equation \ref{eq:PD}), which involves calculating the percent error map between the model and true density distributions. In this appendix, we present for visualization purposes the density distributions. For each of the three synthetic clusters, we present the $N_{\rm S}=100$ {\tt HoliGRALE} and {\tt GRALE} $\kappa$ maps in Figure \ref{fig:massreconmaps}.

\section{Appendix B: Resolution Dependence of Cross-Correlation Functions}\label{txt:crosscorr50}

In Section \ref{txt:correlations}, we computed the cross-correlation function $\omega\left(|\boldsymbol{\theta}|\right)$ (equation \ref{eq:correlationfunc}) to measure how correlated the input image number density is with the percent error (PE$(\boldsymbol{\theta})$). The interpretation of the exercise is that when $\omega\left(|\boldsymbol{\theta}|\right)>0$, the input image number density is anti-correlated with PE$(\boldsymbol{\theta})$, indicating that the model is more accurately reconstructing the density distribution in regions with a higher image number density on distance scales of $|\boldsymbol{\theta}|$. From the test, we were able to conclude that the accuracy of {\tt HoliGRALE} is greater in regions with high image number density regardless of the cluster complexity. Furthermore, we concluded that {\tt HoliGRALE} should give more reliable constraints near cluster member galaxies.

\begin{figure}
\includegraphics[trim={0cm 0cm 0cm 0cm},clip,width=0.33\textwidth]{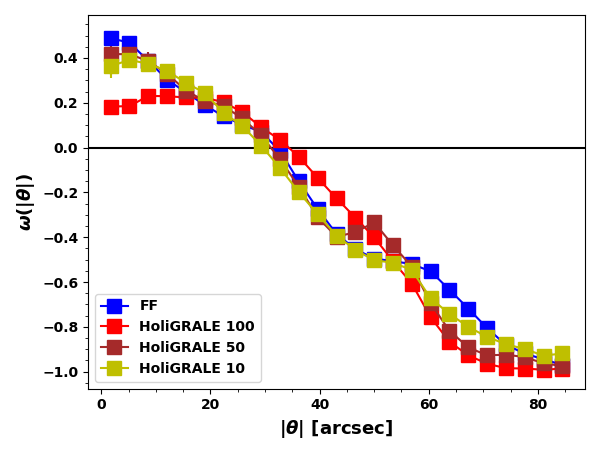}
\includegraphics[trim={0cm 0cm 0cm 0cm},clip,width=0.33\textwidth]{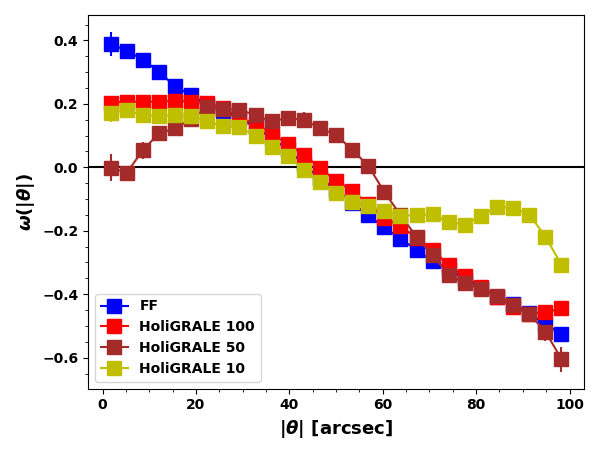}
\includegraphics[trim={0cm 0cm 0cm 0cm},clip,width=0.33\textwidth]{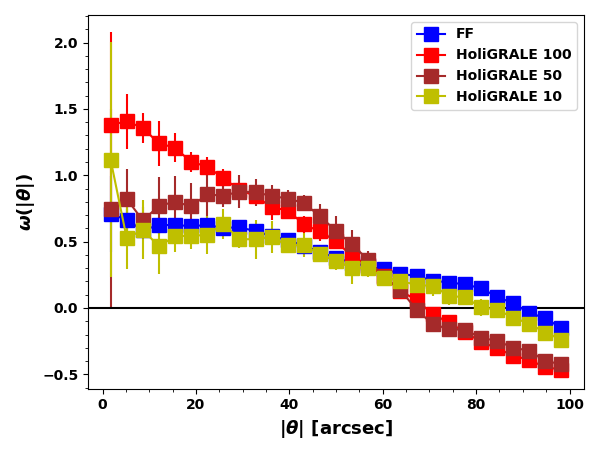}
\caption{Same as Figure \ref{fig:crosscorr} but for a lower resolution of 50$\times$50 pixels for Harriet (\textit{Left}), Maka Ska (\textit{Middle}), and Nokomis (\textit{Right}).}
\label{fig:crosscorr50}
\end{figure}

An important consideration with $\omega\left(|\boldsymbol{\theta}|\right)$ is that it is resolution dependent. In Section \ref{txt:correlations}, we conducted the test on a 100$\times$100 pixel grid on the same window of Figure \ref{fig:synthclusters}. Here, we repeat the same exercise but on a coarser resolution grid of 50$\times$50 pixels in order to examine the resolution dependence. We expect for a coarser resolution grid that the accuracy of all models should be more correlated with image number density, since small scale structure in the models are smoothed out. Our results are shown in Figure \ref{fig:crosscorr50}. As can be seen, in all three synthetic clusters the relations are similar to one another, consistent with more accurate reconstructions in higher image number density regions.

Comparing these results for Maka Ska and Nokomis with our original 100$\times$100 pixel test (see Figure \ref{fig:crosscorr}) and we see that the lack of dependence on image number density observed in {\tt GRALE} is gone for the coarser resolution 50$\times$50 pixel grid. This uncovers the reason for the poor correlation of image number density and PE$(\boldsymbol{\theta})$, as on a better resolution grid where small-scale structure in the density remains, freeform {\tt GRALE} is dominated by noise. On a coarser resolution grid, where small-scale structure is smoothed out, freeform {\tt GRALE} exhibits the trend of image number density and PE$(\boldsymbol{\theta})$. Since {\tt HoliGRALE} is resistant to the change in resolution, we conclude that {\tt HoliGRALE} is able to more reliably reconstruct small-scale structure, particularly near cluster members, than {\tt GRALE}.

\section{Appendix C: Magnification and Time Delay Reconstruction for Synthetic Clusters}\label{txt:mutdtab}

In Section \ref{txt:mutdrecon}, we evaluate the magnification and time delay reconstructions for the three synthetic clusters. Here, we present Table \ref{tab:mutdresid} which shows all our results for each model.

\begin{table*}
\centering
\caption{Magnification and Time Delay Reconstruction with Synthetic Clusters}
\label{tab:mutdresid}
	\begin{tabular}{ccccccc} 
		\hline
		& $N_{\rm S}$ & RMSD$(\mu)$ & RMSD$(|\mu|>5)$ & RMSD$(\Delta\tau)$  & RMSD$(\Delta\tau<5000)$  \\
		\hline
  		& 10 & 8.08 & N/A & 0.15 & 0.33 \\
  		& 20 & 0.54 & 0.16 & 0.01 & N/A \\
  		& 30 & 3.54 & 0.13 & 0.17 & 0.32 \\
  		& 40 & 1.27 & 0.32 & 2.11 & 5.61 \\
  		& 50 & 0.78 & 0.3 & 0.44 & 0.87 \\
  		Harriet & 60 & 0.71 & 0.12 & 0.14 & 0.21 \\
  	    & 70 & 3.63 & 0.57 & 0.04 & 0.10 \\
  		& 80 & 0.44 & 0.23 & 0.09 & 0.33  \\
  		& 90 & 0.19 & 0.15 & 0.14 & 0.37 \\
  		& 100 & 0.74 & 0.16 & 0.56 & 1.15 \\
  		& 100 (FF) & 2.12 & 0.27 & 0.23 & 0.52 \\
		\hline
  		& 10 & 8.02 & 0.18 & 0.59 & 1.16 \\
  		& 20 & 32.15 & 2.77 & 0.09 & 0.25 \\
  		& 30 & 3.70 & 2.63 & 1.18 & 3.25 \\
  		& 40 & 3.38 & 0.93 & 0.10 & 0.26  \\
  		& 50 & 7.51 & 2.05 & 1.29 & 6.76  \\
  		Maka Ska & 60 & 1.46 & 0.44 & 0.13 & 0.43  \\
  	    & 70 & 1.09 & 0.73 & 1.70 & 5.52  \\
  		& 80 & 3.01 & 1.06 & 67.80 & 222.43 \\
  		& 90 & 7.52 & 20.08 & 0.48 & 1.06 \\
  		& 100 & 1.33 & 0.95 & 1.24 & 4.06  \\
  		& 100 (FF) & 4.06 & 2.22 & 4.33 & 14.23 \\
		\hline
  		& 10 & 12.16 & 4.08 & 0.13 & 0.33 \\
  		& 20 & 4.65 & 0.24 & 0.54 & 1.87 \\
  		& 30 & 4.48 & 0.20 & 0.20 & 0.69 \\
  		& 40 & 1.50 & 3.49 & 0.90 & 3.14 \\
  		& 50 & 3.39 & 0.50 & 0.29 & 1.03 \\
  		Nokomis & 60 & 0.52 & 0.32 & 0.15 & 0.52 \\
  	    & 70 & 16.63 & 1.70 & 4.13 & 17.27 \\
  		& 80 & 5.28 & 0.48 & 0.46 & 2.12  \\
  		& 90 & 3.22 & 0.26 & 0.25 & 0.87 \\
  		& 100 & 3.17 & 0.34 & 0.62 & 1.96 \\
  		& 100 (FF) & 4.29 & 1.20 & 0.77 & 2.41 \\
		\hline
	\end{tabular}\\
\medskip{Magnification and time delay reconstruction results for the three synthetic clusters evaluated in this work. RMSD values (equation \ref{eq:rmsd}) are computed for the whole image samples and for a subset. We consider subsets of images with $|\mu|>5$ and $\Delta\tau<5000$ as these populations are the primary of interest images typically observed. }
\end{table*}

\section{Appendix D: SN Requiem and Encore Time Delay Predictions}\label{txt:encoreTDpreds}

Here we present the predicted time delays for future arriving images of SN\,Requiem (Table \ref{tab:SNRequiemTDpred}) and Encore (Table \ref{tab:SNEncoreTDpred}). We discuss these results in more detail in Section \ref{txt:tdpred}.

\begin{table*}[htp]
\centering
\caption{SN\,Requiem Time Delay Predictions}
\label{tab:SNRequiemTDpred}
\setlength{\tabcolsep}{6pt}
\begin{tabular}{@{}cccccc@{}}
\hline
& FF & HG\_noTD\_SP & HG\_noTD\_MP & HG\_wTD\_SP & HG\_wTD\_MP \\
\hline
Bootstrapped $\Delta\tau_{\rm 2d,2a}$ & $1238.82^{+2445.75}_{-1153.42}$ & $4551.67^{+45.04}_{-58.09}$ & $4469.39^{+35.54}_{-39.37}$ & $4662.56^{+29.12}_{-29.97}$ & $4764.77^{+22.96}_{-26.56}$ \\
Ensemble Averaged $\Delta\tau_{\rm 2d,2a}$ & $3276.59^{+1274.56}_{-2696.12}$ & $3995.22^{+833.47}_{-2568.18}$ & $4121.5^{+626.17}_{-1101.44}$ & $4500.57^{+358.73}_{-581.85}$ & $4346.61^{+490.27}_{-818.36}$ \\
Bootstrapped $\Delta\tau_{\rm 2e,2a}$ & $1238.82^{+2445.75}_{-1153.42}$ & $4568.57^{+43.33}_{-61.80}$ & $4488.76^{+38.5}_{-41.63}$ & $4677.07^{+30.09}_{-29.70}$ & $4781.32^{+23.12}_{-25.55}$ \\
Ensemble Averaged $\Delta\tau_{\rm 2e,2a}$ & $3280.82^{+1277.08}_{-2699.52}$ & $4020.7^{+835.47}_{-2577.24}$ & $4185.3^{+619.93}_{-1129.15}$ & $4515.04^{+360.23}_{-581.98}$ & $4400.65^{+442.39}_{-865.22}$ \\
\hline
\end{tabular}\\
\medskip{Time Delay predictions for the 4th arriving (2d) and 5th arriving (2e) images of SN\,Requiem for our 5 lens models.}
\end{table*}

\begin{table*}[htp]
\centering
\caption{SN\,Encore Time Delay Predictions}
\label{tab:SNEncoreTDpred}
\setlength{\tabcolsep}{6pt}
\begin{tabular}{@{}cccccc@{}}
\hline
& FF & HG\_noTD\_SP & HG\_noTD\_MP & HG\_wTD\_SP & HG\_wTD\_MP \\
\hline
Bootstrapped $\Delta\tau_{\rm 1c,1a}$ & $-307.36^{+42.84}_{-34.23}$ & $-323.46^{+220.48}_{-74.58}$ & $-363.95^{+68.67}_{-47.08}$ & $-462.64^{+7.01}_{-8.74}$ & $-463.24^{+10.13}_{-12.11}$ \\
Ensemble Averaged $\Delta\tau_{\rm 1c,1a}$ & $-556.11^{+176.36}_{-217.05}$ & $-392.49^{+100.89}_{-113.99}$ & $-381.67^{+81.12}_{-84.5}$ & $-437.82^{+44.55}_{-36.15}$ & $-442.45^{+65.39}_{-58.59}$ \\
Bootstrapped $\Delta\tau_{\rm 1d,1a}$ & $3889.5^{+53.99}_{-54.86}$ & $4016.76^{+257.01}_{-95.30}$ & $3889.79^{+84.15}_{-61.70}$ & $4022.11^{+25.90}_{-32.43}$ & $4062.46^{+22.54}_{-42.07}$ \\
Ensemble Averaged $\Delta\tau_{\rm 1d,1a}$ & $3262.44^{+700.97}_{-1223.87}$ & $3768.67^{+437.14}_{-532.65}$ & $3761.49^{+204.01}_{-205.8}$ & $4013.46^{+181.89}_{-155.31}$ & $4064.44^{+154.2}_{-204.92}$ \\
Bootstrapped $\Delta\tau_{\rm 1e,1a}$ & $3937.26^{+52.52}_{-56.88}$ & $4058.57^{+233.7}_{-86.29}$ & $3921.12^{+80.48}_{-56.18}$ & $4012.03^{+20.5}_{-17.2}$ & $4066.8^{+17.51}_{-21.07}$ \\
Ensemble Averaged $\Delta\tau_{\rm 1e,1a}$ & $3362.69^{+626.25}_{-1247.57}$ & $3825.8^{+340.53}_{-522.38}$ & $3842.42^{+157.49}_{-154.51}$ & $4028.65^{+171.97}_{-147.08}$ & $4093.51^{+127.44}_{-152.32}$ \\
\hline
\end{tabular}\\
\medskip{Time Delay predictions for the 1st arriving (1c), 4th arriving (1d) and 5th arriving (1e) images of SN\,Encore for our 5 lens models.}
\end{table*}

\label{lastpage}
\end{document}